\documentclass[aip,twocolumn,letterpaper]{revtex4}

\usepackage{fullpage}

\usepackage{xcolor}

\usepackage{graphics}
\usepackage{epsfig}

\begin{document}
\title{Why carbon dioxide makes stellarators so important}
\author{Allen H. Boozer}
\affiliation{Columbia University, New York, NY  10027\\ ahb17@columbia.edu}

\begin{abstract}

The increasing level of atmospheric carbon dioxide has driven public discourse throughout the world.  An immediate implementation of carbon-free energy sources is demanded with little discussion of  costs,  technical constraints on the sources, or  implications of high residual levels of carbon dioxide.  Residual carbon-dioxide can be removed from the air, but the cost to remove the carbon-dioxide produced by human activity during a year is thought to be trillions of dollars---otherwise it remains in the atmosphere for centuries.   Economic considerations may limit wind and solar sources to less than 40\% of the electricity production.   Fission or fusion may be the only choice for most of the rest.  Development costs are orders of magnitude smaller than implementation costs, which are tens of trillions of dollars for fission.  A needless delay in the development of fusion has enormous financial implications.  As will be shown stellarators are better positioned  than any other concept for a fast path to fusion.   A computationally derived conceptual design for a stellarator reactor may allow final design and construction to be initiated without the delay of intermediate generations of experiments.   The most urgent issue is the development of conceptual designs. 

\end{abstract}

\date{\today} 
\maketitle




\section{Introduction}

Societal risks associated with the increase in atmospheric carbon dioxide make the rapid development of fusion energy compelling.  This is emphasized by a 2019 report of the Organization for Economic Co-operation and Development (OECD)  \cite{wind-solar fraction}, which found a large cost penalty when wind and solar exceed 30\% to 40\% of the electricity-generation fraction.  Of all fusion concepts the stellarator appears best poised for rapid development.  

 The increase in atmospheric carbon-dioxide is unleashing powerful political forces but has aroused  little interest in determining the options that science could offer---neither carbon-dioxide removal nor carbon-free energy sources, such as fusion.   As will be shown, the development of options costs several orders of magnitude less than their deployment.  This enormous cost ratio makes it irrational to implement carbon-dioxide mitigation without also having research focused on the fastest possible development of better options.
 
The risks associated with the increase in carbon-dioxide are sometimes described in apocalyptic terms---neither limitations of finance nor on governmental power should be allowed to stand in the way of implementing emergency measures.   The term ``moral hazard" means the encouragement of risk over responsibility.  The development of science-based options can be viewed as a moral hazard by removing the mandate for immediate implementation. 

Nevertheless, ill-considered options can entail enormous financial costs while exposing the world to the risks of elevated carbon-dioxide levels.  Such options can have a large moral hazard by blocking the rapid development of more effective options.  

The prominently discussed options are not solutions in the sense of returning carbon-dioxide to a pre-industrial-revolution level.  They seek to only limit the carbon-dioxide increase to a tolerable level.  A tolerable level \cite{NAS-CO2} is often taken to be consistent with a temperature increase of $2^o$~C.  But, the temperature increase that will occur under various scenarios is uncertain as are the effects, detrimental and beneficial, that arise at various carbon-dioxide levels.  The options for a rapid transition to carbon-free electricity production frequently ignore the OECD limit \cite{wind-solar fraction} on the fraction of the electricity-generation that can be economically produced by solar and wind, 30\% to 40\%.  Whatever strategy is adopted, it should be consistent with an increasing use of energy.  For example, Table 49 in \cite{UN:energy} shows an approximate 4\% annual increase in the electricity-generation capacity of the world.

When elevated levels of atmospheric carbon dioxide are perceived to have dangers, the cost of removal of  the carbon dioxide that humans place in the atmosphere \cite{NAS-CO2} during a year, $\sim 50$~Gt, defines the financial risk of not moving to carbon-free energy sources.    As discussed in Chapter 1 of  \cite{NAS-CO2}, the natural removal of carbon-dioxide from the atmosphere requires centuries \cite{NAS-CO2}.   

The cost of removal is not the standard method of estimating the cost of carbon-dioxide emissions.  The standard method is the Social Cost of Carbon (SCC), which uses models to estimate financial implications: damages and benefits.  A recent review of the literature \cite{SCC} finds the mean value of the estimates of the social cost of carbon-dioxide emission is $\sim$\$50/t.  When a discount rate is included that cost is reduced to $\sim$\$30/t.  The range of estimates is large, but using the \$30/t estimate, the damage produced by emitting 50~Gt is approximately \$1.5 trillion.  The OECD report \cite{wind-solar fraction} discussed carbon-dioxide pricing at \$50/t as a policy instrument to internalize costs and achieve a low emission outcome.  The cost of 50~Gt at \$50/t is \$2.5 trillion.

Removal is the only ensured way to avoid effects of carbon-dioxide emission.  Direct air capture facilities could remove a year's emission quickly but at an annual operating cost that is projected  \cite{NAS-CO2} to lie between \$5~trillion and \$30~trillion.   The capital cost, which was estimated in 2019 by Fasihi et al \cite{Fasihi:2019}, also has large uncertainties, but \$200 to \$300 per t/yr may be credible.  This estimate gives a cost of  \$10 to \$15 trillion for facilities that would remove 50~Gt/yr.  If the capital costs were sufficiently low, these facilities could be operated intermittently using extremely low-cost wind and solar power.

Costs should be compared to the world economic output,  which is estimated  \cite{IMF} to be  \$92~trillion in 2020, and the wholesale value of annual electricity production, approximately \$2 trillion a year.  This can be derived by multiplying the \$80/MWh for the wholesale price of electricity in Table 8 on p. 127 of \cite{wind-solar fraction} by the 2016 world electricity production of 25$\times10^9$~MWh given in Table 37 of \cite{UN:energy}.  

The costs of carbon-dioxide removal have large uncertainties, and the research proposed in the 2019 National Academy study \cite{NAS-CO2}, with a maximal annual expenditure of less than \$250 million, is clearly not designed to develop minimal cost options on the shortest possible time scale.  Chapter five of the study \cite{NAS-CO2} said a facility  that would remove $10^4$~t/yr of carbon-dioxide could demonstrate the technology and would cost approximately \$100 million.  The ratio of the cost of deployment to the cost of development of a carbon-dioxide removal option is measured by the ratio of the \$10 trillion estimated as the capital cost to the \$100 million required for a demonstration facility, a factor of a hundred thousand.

The time required to develop science-based options can be studied but not definitively answered without their development.  Nevertheless, science can move with remarkable speed in periods of societal crisis. The splitting of the nucleus in December 1938 to the launching of the first nuclear-powered submarine in January 1954 was approximately fifteen years; nuclear weapon development required less than seven.  The Apollo program to land and return a person from the moon  was announced in May 1961 and reached its goal in July 1969, just over eight years.  The time commonly envisioned for addressing the carbon-dioxide problem is thirty years \cite{wind-solar fraction,NAS-CO2}, which includes the time for deployment as well as for the development of a solution. 


What are the options for carbon-free energy production?  Energy from nuclear fission is carbon free, but has proliferation, radioactive waste, and safety issues.  Wind and solar provide cost-effective carbon-free options, but both suffer from site specificity, intermittency, and grid stability issues.  Intermittency accounts in part for the discrepancy between wind and solar providing 32\% of the electricity generation capacity in 2016 but only 25\% of the electricity \cite{UN:energy}.  The intermittency issue can be addressed by batteries, which must be large and expensive when wind and solar dominate a grid, and by power transmission over a scale greater than that of weather patterns.  A more subtle issue is the stability of the electrical grid.  Wind and solar power eliminate the heavy rotating generators of conventional power plants, which have sufficient inertia to stabilize the grid during changing power loads  \cite{grid-stability}.   Chapter 5 of the 2019 OECD report \cite{wind-solar fraction} stated that  financial feasibility limits the electricity-generation fraction that can be produced by solar and wind to 30\% to 40\%.   

The cost of deploying carbon-free energy sources can be assessed by multiplying the overnight costs of electricity-generation technologies by the electricity generation capacity of the world, which is approximately 8000~GW in 2020 (Table 49 in \cite{UN:energy}).  The overnight cost is the expense of constructing a generation facility, ignoring the interest charges, and dividing by the number of Watts generated.  The overnight cost \cite{overnight} is \$1.32/W for wind, \$1.33/W for tracking solar photovoltaic, and \$6.32/W for advanced nuclear fission.  Replacing the present electric generating capacity with new nuclear fission facilities would cost approximately \$50~trillion.  Replacement with wind and solar facilities would cost approximately \$10~trillion, but this cost contains no provision for addressing intermittency and grid-stability issues.  Wind and solar appear to be obvious choices to replace some fraction of the electricity generation, but that fraction may be very limited \cite{wind-solar fraction} and is an important topic for additional research.

Fusion energy has fundamental advantages compared to alternative carbon-free energy sources---especially when most of the electricity-generation capacity should come from sources other than wind or solar.  The development costs of fusion energy are low compared to deployment costs.  The ratio of deployment to development costs is largely determined by the ratio of the 8000~GW of total generation capacity to the envisioned size of a demonstration magnetic fusion reactor, approximately 1~GW. 

The low cost of developing a fusion option, compared to the societal risks of not doing so, makes an assessment of rapid paths to fusion development imperative.  A needless delay imposes enormous costs on society: the cost of carbon-dioxide removal---trillions of dollars per year of delay---and the cost of a less than optimal replacement for a large fraction of the electricity-generation capacity---ten of trillions of dollars spread over a few decades.

When societal risks are considered, the case for stellarators is compelling.  The stellarator, among all fusion concepts, has properties that best open a fast and low-risk path to reactors. The plasma in a stellarator is externally controlled, rather than self-organized, to a far greater extent than in any other fusion concept---magnetic or inertial.  In addition, the stellarator can make use of an order of magnitude more distributions of external magnetic fields.  These differences lead to two distinct research paradigms: optimization using computational design confirmed by experiments for stellarators and extrapolation from one generation of experiments to another as in tokamaks.  

Section \ref{sec:path} compares stellarators and tokamaks, which clarifies why with present knowledge the stellarator appears to offer a far better path for a fast and low-risk development of a fusion reactor.  The time required for development is essentially determined by the number of consecutive generations of experiments that are needed. Relative risk is determined by the issues requiring proof-of-principle demonstration, such as disruption avoidance in tokamaks.  Using these criteria, the step to a power plant from the ITER tokamak appears more difficult than going directly using our present understanding of stellarators.  Tokamaks are a certain geometric limit of quasi-axisymmetric stellarators, but with only 10\% of the possible external magnetic field distributions available to them.   It should not be surprising that an additional order of magnitude in design freedom aids the achievement of fusion energy.

Agreement with the statement that stellarators offer a far more likely path for a fast and low-risk development of a fusion reactor does not imply ITER or the tokamak should be terminated.  Deuterium-tritium experiments on ITER, which are scheduled to begin in fifteen years, followed by a tokamak demonstration reactor may provide a longterm option for carbon-free energy, and that may prove to be important.  But, an ITER-centric time scale may not be the fastest possible path to a demonstration of fusion energy.  It was only fifteen years between the splitting of the uranium nucleus and a fission-powered submarine.

A necessary step in stellarator development is the formulation of conceptual designs for one or more attractive, low-risk, stellarator demonstration reactors.  Major improvements in stellarator reactors can be made through computational design, which is fast and has a low cost.  No reason is known why a conceptual design for a stellarator reactor cannot be sufficiently attractive that its final design and construction could be initiated without the delay of intermediate generations of experiments.  Clarifying and confirming experiments that are constructed simultaneously do not produce delays.  Many such experiments were built while TFTR and JET were under construction, such as the tokamak now called DIII-D.

In 2018 the U.S. stellarator community published an article \emph{Stellarator Research Opportunities} \cite{NSCC:2018}, which reviewed the issues of stellarators and should be consulted for additional details and references.  The \emph{Stellarator Research Opportunities} article also defined a research program, but that program did not consider the implications of societal risks.  

Here societal risks are the focus, which makes the primary question whether the development time for fusion power could be shortened by eliminating a generation or more of major stellarator experiments before beginning construction of a reactor.  A major experiment means with a time scale of a decade or longer.  In a development program defined by its urgency, that question may be best answered as part of the review process of conceptual designs for stellarator reactors.  The question addressed in this paper is the urgency of the development of conceptual designs, which has a low total cost---presumably well under \$100 million---compared to the overall cost of developing stellarator reactors.

The sections following Section \ref{sec:path} provide additional details on important topics and areas in which major improvements can be quickly made through computational design.  Section \ref{sec:coils} discusses coil issues including the importance of open access, which has had almost no mention in the fusion literature, possibly because the type of open access available in stellarators does not appear to be possible in tokamaks.  Section \ref{sec:configurations} discusses the space of stellarator configurations, which is so large that special strategies are required to determine those most suitable for fusion reactors.  Section \ref{sec:turbulence} discusses strategies for dealing with the implications of microturbulent transport on reactor design.  A method of assessing the implications is developed in the Appendix.  Appendix \ref{sec:transport model} shows that gyo-Bohm scaling fits the overall scaling laws of both tokamaks and stellarators with remarkable accuracy and derives implications for reactor design.  Section \ref{sec:edge} considers divertors and the protection of the walls from alpha-particle damage in stellarators.  Section \ref{sec:technical} discusses technical developments in the areas of coils, liquid films for the walls, solid walls, and breeding blankets.


\section{Stellarators as a path for the rapid development of fusion \label{sec:path} }

The stellarator, among all fusion concepts, has properties that best open a fast and reliable path to reactors.  These properties can be illustrated by comparing stellarators with tokamaks.  Far more tokamak than stellarator experiments have been performed, but far more details about fusion plasmas are required for the design of a tokamak than of a stellarator reactor.  

The extra information required for tokamak reactors makes the step from ITER to a demonstration power plant (DEMO) appear more difficult than going from our present understanding of stellarators to a stellarator DEMO.  Open questions that ITER will address are summarized in a 2019 article by Hawryluk and Zohm in \emph{Physics Today}  \cite{Physics Today:2019} and the issues that are being considered for the European tokamak demonstration power plant are reviewed in a 2019 article in \emph{Nuclear Fusion} \cite{EU-DEMO2019}.

\begin{enumerate}

\item  \textbf{No proof-of-principle issue, such as disruption avoidance in tokamaks, blocks rapid development of stellarators.}

Disruptions are an existential threat to reactor-scale tokamaks, particularly the threat of strong currents of relativistic electrons.  Nevertheless, disruptions receive only cursory consideration in the papers on open questions that will be addressed by ITER \cite{Physics Today:2019} and on the European tokamak demonstration reactor \cite{EU-DEMO2019}.   

As discussed below, no solution that is generally perceived to be reliable is known for disruptions, which makes all tokamak planning problematic.  A common presumption is that the disruption problem will be solved because it must.

\item  \textbf{The stellarator is unique among all fusion concepts, magnetic and inertial, in not using the plasma itself to provide an essential part of its confinement concept.} \vspace{0.1in}

This allows stellarators to be designed computationally with far more reliability than any other fusion concept.  

The alternative to computational design is extrapolation from one generation of experiments to another, as is traditional in tokamaks.  The abstract of the original paper on the ITER Physics Basis emphasized extrapolation \cite{ITER physics}.  The paper that introduced the scientific basis of W7-X emphasized computational optimization of designs \cite{Grieger:1992}.   

Four disadvantages of extrapolation in comparison to computational design are:
\begin{enumerate}

\item Experiments build in conservatism---even apparently minor changes in design are not possible and therefore remain unstudied.

\item  Experiments are built and operated over long periods of time---often a number of decades.  \vspace{0.1in}

Multiple experiments carried out at the same time do not delay development, but extrapolation using consecutive generations of experiments does.  The need for consecutive generations of experiments should be minimized.

\item  The cost of computational design is many orders of magnitude smaller than building a major experiment, as well as having a much faster time scale. 

\item  Extrapolations are dangerous when changing physics regimes.  Examples are (i) plasma control in ignited versus non-ignited plasmas and (ii) the formation of a current of relativistic electrons during a disruption. 

In existing tokamaks, external heating provides plasma control that is not present when the heating is dominated by DT fusion.  As will be discussed, a stellarator has far more available degrees of freedom for control but requires fewer than a tokamak. 

Although relativistic electrons are an existential threat to tokamak reactors, the danger is removed in standard stellarator designs.  In stellarators, the plasma is robustly centered within the confinement chamber, and the effect of plasma currents on the rotational transform, which is the derivative of the poloidal relative to the toroidal flux, is minimized.  The loop voltage, which accelerates electrons, is the rate of slippage of the poloidal relative to the toroidal  magnetic flux.   

In tokamaks, a loss of position control accompanies disruptions and each megaampere of decay in the plasma current can increase the current in relativistic electrons by a factor of ten.  As noted in Chapter 3, Table 5, of the ITER physics basis \cite{Hender:2007}, this implies an amplification of a current of relativistic electrons a trillion times greater in ITER than in JET.   Recent theoretical work \cite{Hesslow:2019} indicates the amplification may be far larger than formerly expected.  

The danger of electrons running away to relativistic energies has been prominent in the literature for more than twenty years, but no method has yet been devised of mitigating the danger in a way that is perceived to be reliable.  The last paragraph of the 2019 \emph{Nuclear Fusion} review of the physics of runaway electrons (RE) \cite{Breizman:2019} noted:  \emph{With ITER construction in progress, reliable means of RE mitigation are yet to be developed.}  Sections 10.8 and 11.3 of this review discussed the wall damage that runaway electrons can produce.  A runaway current as small as 300 kA could cause the melting limits of a wall panel roof to be exceeded.  This is a worse-case number, and several megaamperes of relativistic electrons striking the walls may be required for extreme damage.

The  severity of the damage that can be produced by even a single relativistic-electron incident implies: (i) The achievement of the ITER mission will be difficult when more than one such incident occurs in a year.  (ii) The strategy for avoidance must be fundamentally based on theory and computation.  

Tokamak disruptions are often said to result from exceeding operating boundaries \cite{Physics Today:2019}.  Unfortunately, methods of steering  tokamak plasmas away from operating boundaries during a fusion burn are extremely limited and slow, seconds for the temperature and density profiles and minutes for the current profile.  

Steering alone does not provide complete protection against disruptions.  A disruption can be initiated if a part of a wall tile or even a tiny flake from the tungsten diverter targets were to enter the plasma \cite{tungsten flakes:2019}.  But, in Alcator C-­Mod disruptions caused by tungsten flakes could be avoided when the poloidal magnetic field could be maintained \cite{tungsten flakes:2019}.   Of course, a tile falling into a stellarator plasma would cause a rapid drop in the plasma pressure.  Stellarators should be designed so currents associated with the diamagnetic effect of the plasma and the quick loss of the plasma energy through radiation would be tolerable.  These requirements are far less demanding than those of a tokamak.

\end{enumerate}

\item  \textbf{Stellarator reactor designs are only weakly dependent on the plasma pressure profile.}

\begin{enumerate}

\item  The sensitivity of tokamaks to the profile of the current density makes them highly sensitive to the pressure profile, through the bootstrap current, and in non-steady-state reactors to  the temperature profile, through the resistivity. 

\item  Microturbulent transport is an issue for all magnetic-fusion systems.  The insensitivity of stellarators to the pressure profile implies that only the overall level of the transport is of central importance.  For tokamaks, not only is the overall level important, but also the radial dependence of the transport.

\item The overall effective transport rate can be normalized to gyro-Bohm transport by a coefficient $\mathcal{D}$,  Appendix \ref{sec:fusion and transport}.   The $H$-factor of the tokamak literature scales as $H\propto 1/\mathcal{D}^{2/5}$. 

Too large a $\mathcal{D}$ implies that either the power output of a single reactor or the magnetic field strength become excessively large.  Too small a $\mathcal{D}$ implies the plasma radius is small compared to the thickness of the blankets and shields, which means the power production is too small compared to the reactor cost.   The problem of too small a $\mathcal{D}$ could be addressed by reducing the plasma reactivity by departing from a 50/50 deuterium-tritium mixture.  

$\mathcal{D}$ of order but somewhat smaller than unity is optimal.  

\item Stellarators do not have constraints such as the Greenwald Limit on the plasma density or the high electron temperature required in tokamaks for current maintenance.  The higher the electron temperature the greater the number of energetic alpha particles, which increases the sensitivity to energetic particle instabilities.  As discussed in Appendix \ref{sec:fusion and transport}, the degradation of confinement with power seen in empirical scaling laws implies a degradation of confinement with temperature.


\end{enumerate}

\item \textbf{Stellarators offer far more freedom of control than do tokamaks.} \hspace{0.1in}  

Approximately fifty externally produced distributions of magnetic field are available for plasma control in stellarators; approximately five are available in axisymmetric tokamaks.  Unlike in stellarators, these require careful time-dependent control.  

The plasma profiles in tokamaks require far more control than in stellarators, but the available degrees of freedom to provide that control are far fewer.

\item \textbf{The coil systems in stellarators, unlike those in tokamaks, can be designed for open access to the plasma chamber,} Section \ref{sec:access}. \vspace{0.1in}

If fusion is to be developed rapidly, a demonstration reactor (DEMO) must be designed to allow first wall components to be changed quickly---too many uncertainties remain in first wall materials \cite{Fusion materials:2017,Wall-materials:2019}, in concepts such as walls being covered by liquids 
\cite{Liquid-walls:2019}, and in blankets for breeding tritium \cite{Abdou:2015} for it to be otherwise.

Open access also shortens maintenance times in operating reactors.

\end{enumerate}

Stellarators do have the disadvantage of a larger aspect ratio, the ratio of the major to the minor radius, $R/a$, of the torus.  The power density on the walls $p_w$, megawatts per square meter, should be as large as is consistent with a reasonable wall lifetime to minimize the cost of fusion per kilowatt hour.  The total power output $P_T$ of a single fusion reactor would ideally be small to maximize flexibility and minimize the capital required for the construction of a single unit.  The obvious relation $P_T \propto (R/a) a^2 p_w$  couples a high power density with a high total power output.  The aspect ratio $R/a$ is determined by the fundamental properties of a fusion concept, and the aspect ratio of stellarators is several times larger than that of tokamaks.  The minor radius $a$ cannot be too small compared to the thickness of blankets and shields around a fusion plasma, but otherwise it is determined by transport.

Appendix \ref{sec:fusion and transport} shows that the empirical energy-confinement scaling of tokamak and stellarator experiments closely match what would be expected if the diffusion coefficient were a factor $\mathcal{D}$ times gyro-Bohm transport.   The required plasma radius squared scales with the quality of confinement $1/\mathcal{D}$, the magnetic field strength $B$, the power density on the walls $p_w$, and the central plasma temperature $T_0$ as 
\begin{equation} 
a^2\propto \frac{\mathcal{D}^{4/5} T_0^{6/5} }{B^{8/5} p_w^{2/5} }.
\end{equation}
For a given quality of confinement, wall loading, and aspect ratio, the total power output $P_T\propto a^2$ can be made smaller by using a larger magnetic field and a lower plasma temperature---as long a $T_0>10$~keV.  The higher central temperature in tokamak reactors offsets the advantage of a smaller aspect ratio for allowing fusion power plants to have a smaller total power output $P_T$.  Tokamak reactor designs often require a significant fraction of this power output be used for maintenance of the plasma current and for control while stellarators do not.   The larger recirculating power fraction is a significant burden on the economic viability of tokamak reactors.

Early operations of  the W7-X stellarator achieved $\mathcal{D}=0.13$ in ten-second steady-state plasma conditions, which yields attractive reactor designs, and $\mathcal{D}=0.05$ during short intervals.  DIII-D has carried out long-pulse tokamak experiments that achieved $\mathcal{D}=0.31$.  Both the stellarator and tokamak results for $\mathcal{D}$ are discussed in Appendix \ref{sec:exp-comparison}.



\section{Coils for stellarator reactors \label{sec:coils} }

\begin{figure}
\centerline{ \includegraphics[width=3.0in]{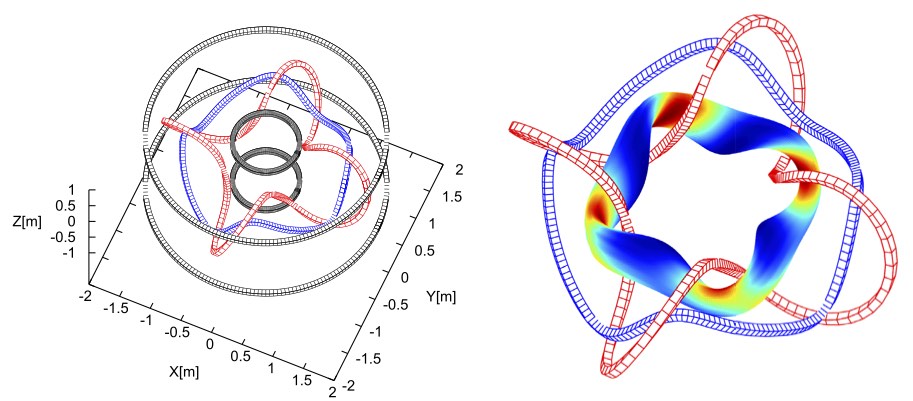} }
\caption{H. Yamaguchi has published  a set of continuous helical coils that generate a magnetic field that approximates a quasi-isodynamic stellarator.  The  full coil set is illustrated on the left and the outer magnetic surface together with the primary coils are illustrated on the right.  As discussed, the red coil is the only coil that need limit access to the plasma chamber.  This is Figure 1 in  H. Yamaguchi,  Nucl. Fusion \textbf{59} 104002 (2019). }
\label{fig:coil_f}
\end{figure}

Three properties of the coils that produce the external magnetic field are of particular importance to a rapid and reliable development of fusion energy:  (1) Coils that offer easy access to the plasma chamber.  (2) Coils that are relatively easy to manufacture because the magnetic fields that they produce are not unnecessarily strong nor rapidly varying in space.  (3) Coil systems that maximize the flexibility of plasma control \cite{Stell-desgn:2015}.

\subsection{ Coils with easy chamber access \label{sec:access} }

Coils systems for both tokamaks and stellarators have been designed in a manner that makes access to the plasma chamber extremely  constrained.   Coils can be designed with demountable joints, so parts of the coils can be removed to provide chamber access \cite{demountable coils}.  For stellarators, but not for tokamaks, a far simpler and more robust method of providing access is available.  The red coil in  Figure \ref{fig:coil_f}, which is from  \cite{Yamaguchi:2019}, is the only coil that need encircle a stellarator plasma and limit plasma access.  Mathematics ensures the rest of the magnetic field could be produced by coils, each shaped like a windowpane, with some embedded in the removable sections of the walls \cite{Stell-desgn:2015}.  There is no necessity for the removal of wall sections to be more restricted than that produced by the red coil  of  Figure \ref{fig:coil_f}.   

Discreteness in the coils that produce the toroidal magnetic field in tokamaks produces an unacceptable toroidal ripple unless the space between toroidal-field coils is small.  Figure \ref{fig:coil_f} shows how  the toroidal ripple can be used in stellarators to provide the helical magnetic field that they require.

The winding and assembly of coils can be simplified if joins are possible during assembly, even if the joints are not demountable.  A demountable coil means parts can be repeatedly separated and rejoined.  The provision of joints is much easier when the joints can be located in a low-field region as they can be in stellarators. 

Rapid changes in the components that surround the plasma are critical for the fast development of fusion energy and would minimize maintenance time in a reactor.  Despite the obvious importance, coil concepts such as the one illustrated in Figure \ref{fig:coil_f} remain largely unexplored.

\subsection{Efficient magnetic field distributions} 

A curl-free magnetic field decays with the distance $x$ from the coil that produces it as $e^{-kx}$ where $k$ is the wavenumber of the field.  All possible external magnetic field distributions can be ordered  by their efficiency of production \cite{Stell-desgn:2015,Landreman:2016}.  The are approximately fifty distributions that have adequate efficiency \cite{Boozer:RMP}.   Stellarator optimizations could be constrained so only magnetic fields that can be produced efficiently at a distance are included, which are the only magnetic fields that can be produced by practical coils.  

The benefits of limiting the design to the efficiently produced external field distributions are largely unexplored.

\subsection{Coils needed for plasma control} 

The important stellarator control parameters are the efficient magnetic field distributions.  It is known that the importance of the various magnetic field distributions for plasma control varies widely \cite{Stell-desgn:2015}.   What has not been done is to assess which of these distributions are the most important and how the control of these distributions can be incorporated in coil design.  

The speed and the completeness with which a given machine allows fusion to be developed is largely determined by its available control.

\section{Stellarator configurations \label{sec:configurations} }

The space in which stellarators are designed has about fifty degrees of freedom---far too many for an optimization code to ensure that a global optimum has been found.   Although a direct and complete optimization is impossible, practical numerical optimizations can (1) refine an initial guess or (2) maintain the optimization of a curl-free magnetic field as the plasma pressure is increased.

\begin{figure*}
\centerline{ \includegraphics[width=6.5in]{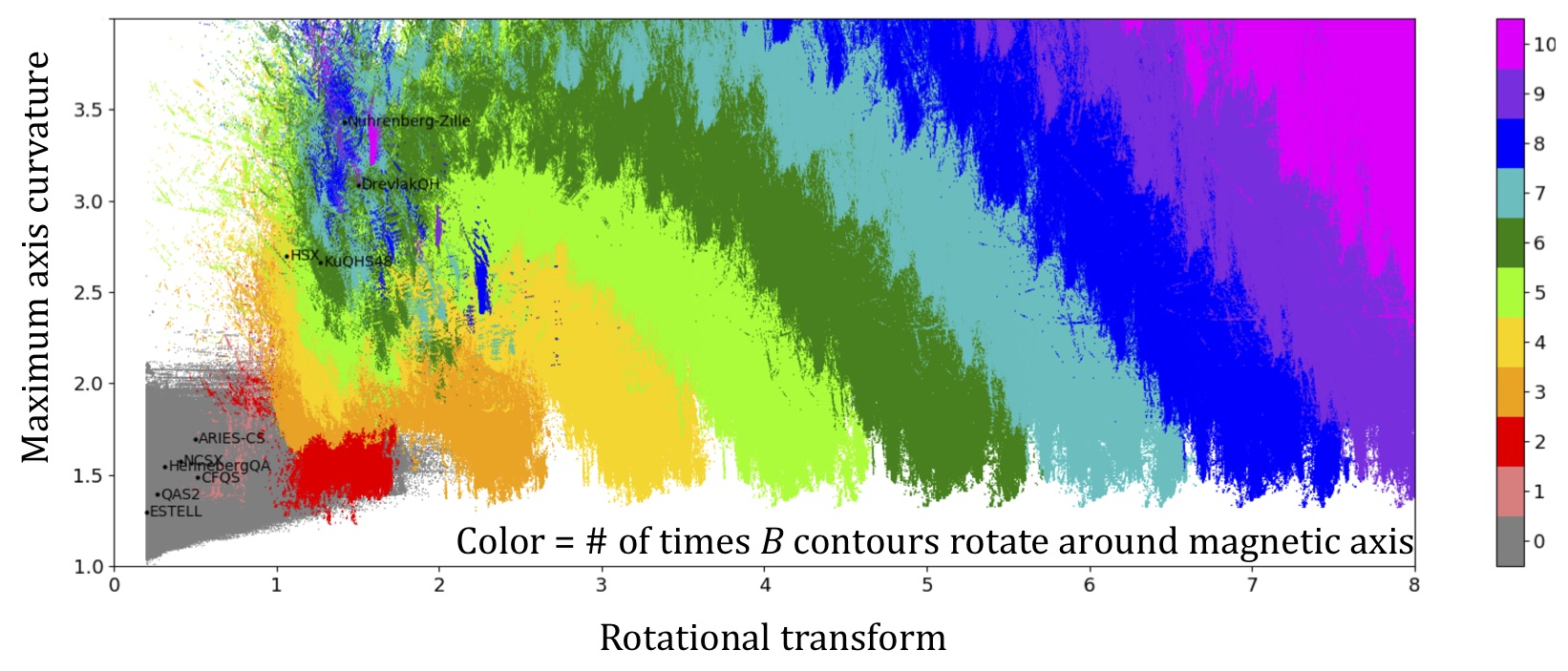} }
\caption{Matt Landreman and his group have used analytic expansions around the magnetic axis to survey the landscape of possible quasisymmetric stellarators.  The figure shows a database of $2.4 \times 10^8$ quasisymmetric stellarator configurations.  The few designated points on the left side of the figure indicate previously known stellarator configurations}
\label{fig:landscape}
\end{figure*}

The large size of this space compared to what has been explored is illustrated by Figure \ref{fig:landscape}.

\subsection{Identification of states for optimization}

The attractiveness of an optimized stellarator is largely determined by the initial state used in the optimization, which makes makes methods of choosing an  initial state of great practical importance.  Two concepts for finding advantageous initial states are (1) a Taylor expansion around the magnetic axis and (2) the optimization of an outer magnetic surface of a curl-free magnetic field.

\subsubsection{Expansion around the axis}

The original idea of defining equilibria using a Taylor expansion around the central field line in a toroidal plasma, the magnetic axis, \cite{Mercier:1964} is due to Mercier in 1964.  Taylor expansion methods were found in 1991 to set important constraints on stellarators \cite{Garren:1991}, and recent advances have been made \cite{Landreman:2019}.  Unlike axisymmetric systems, the achievement of adequate particle confinement is the primary physics issue in stellarators.  Methods of achieving particle confinement, quasisymmetry and omnigenity, are discussed in \cite{Boozer:1983,Cary:1997,Landreman:2012}.


\subsubsection{Optimization of an outer surface}

An outer magnetic surface of a curl-free magnetic field can be found  \cite{stell-surf-Boozer2019} that has desirable confinement properties such as exact quasisymmetry.  Quasisymmetry gives tokamak-like confinement of individual particles.   This method is particularly important in conjunction with the concept of annular design, which is discussed in Section \ref{sec:annular design} and in Appendix  \ref{sec:annulus}. 

The shape of the outer magnetic surface is determined by three functions of two angles that must satisfy constraints.  Using $(R,\zeta,Z)$ cylindrical coordinates, the three functions are $R(\theta,\varphi)$, $\zeta=\varphi+\omega(\theta,\varphi)$, and $Z(\theta,\varphi)$, where $\theta$ and $\varphi$ are the poloidal and toroidal angles in Boozer magnetic coordinates in which the constraint of exact quasi-symmetry is easily specified \cite{Boozer:1983}.   One function of the two angles, which can be $\omega(\theta,\varphi)$, is required to obtain magnetic coordinates leaving $R(\theta,\varphi)$ and $Z(\theta,\varphi)$ free.  Obtaining well confined particle trajectories constrains half of the freedom of another function of the two angles.  For example, quasiaxisymmetry is obtained when the field strength $B$ has the property that $B(\theta,\varphi) = \oint B(\theta,\varphi) d\varphi/2\pi$.  The curl-free solution can be extended throughout the volume enclosed by the surface by choosing efficient magnetic field distributions so the magnetic field perpendicular to the optimization surface is zero.  Maximizing the coil efficiency is equivalent to placing another constraint on a function of the two angles.  There are only two-and-a-half functions of constraints on the three functions of $\theta$ and $\varphi$.  Consequently, there is additional freedom in the properties of the magnetic field.

This method of defining curl-free states for optimization is unexplored.


\subsection{Annular Design \label{sec:annular design}}

An optimal design for a stellarator may have low plasma transport in the outer half of the minor radius, but such rapid transport in the inner half that the pressure is essentially constant there, Appendix \ref{sec:annulus}.  The implications are essentially unexplored, but there are advantages to having the confinement produced by an outer annulus. (1) A spatially constant pressure $p$ maximizes $\int p^2d^3x$  for a fixed maximum pressure, which maximizes the fusion power.  The confinement time of the plasma is the ratio of the total plasma volume to the volume of the annulus longer than the confinement time of the annulus.  (2) Impurities tend to be flushed out more readily the narrower the confinement annulus compared to the total confining volume.  (3) The injection of fuel is easier.  (4)  The optimal 50/50 DT ratio can be maintained in the fusing plasma, which is not trivial when transport coefficients are small in core. (5) Control of the width of the annulus would provide an important control of the plasma.

The situation in tokamaks with good confinement is related but different.  The pressure drops across the core of a tokamak, but there is a narrow region right at the plasma edge, where a transport barrier arises that creates a pedestal, which raises the plasma pressure everywhere inside.  See Figure 2 in \cite{Physics Today:2019} and the related discussion.  This annulus naturally has periodic instabilities called Edge Localized Modes (ELM's), which must be mitigated to avoid unacceptable damage to the chamber walls, \cite{Hender:2007}.  The extent to which such pedestals arise in stellarators, or whether it is even desirable that they occur, is unclear.  The self-organized state of an axisymmetric tokamak plasma implies the control over important features such as the pressure profile is limited.  Carefully designed non-axisymmetric perturbations that preserve the quasisymmetry of the tokamak core could ameliorate this limitation \cite{Boozer:Nature2018}.  An example  is the control of ELM's by long wavelength non-axisymmetric magnetic fields \cite{Park:2018}.


\section{Microturbulence strategies \label{sec:turbulence} }

An uncertainty in the design of both stellarator and tokamak fusion reactors is microturbulent transport.  The physics of microturbulence in stellarators was reviewed in 2015 by Helander et al \cite{Helander:2015}.  The two most important types of microturbulence are the ion-temperature-gradient (ITG) mode and the trapped electron (TE) mode. The TE instability has much greater stability when the trapped electrons are primarily in a region of good magnetic field line curvature as in W7-X.  In tokamaks and in quasi-axisymmetric and quasi-helically symmetric stellarators the trapped electrons are primarily in a region of bad curvature. 

ITG microturbulence appears to have a beneficial effect of expelling impurities, which implies some level is desirable.  But, in a reactor ITG turbulence can not be so large that it unacceptably degrades ion confinement.  Unlike the situation in tokamaks, the details of the pressure profile in stellarators are of little relevance.  ITG turbulence need only be kept at a level that is consistent with an adequate fusion product, $n\tau_ET$, where $n$ is the number density of the deuterium and tritium ions, $\tau_E$ is the energy confinement time, and $T$ is the temperature.

As discussed in Appendix \ref{sec:fusion and transport}, power-law scaling relations hold with remarkable accuracy for tokamaks and stellarators.  Nevertheless, scaling relations do not provide the certainty that is wanted for a reactor design---even in stellarators.  The effect of microturbulence is not well understood in either tokamaks or stellarators.  Non-linear calculations of microturbulence using the GENE code \cite{Helander:2015} show a W7-X case with a transport enhancement of twenty times the characteristic gyro-Bohm value and a DIII-D case with an enhancement of two-hundred times.


In designing a stellarator reactor, the most important information on microturbulence is what factors are beneficial in obtaining an adequate  $n \tau_E T$.  A higher magnetic field strength $B$ appears to be clearly beneficial.  The ion temperature gradient, $d\ln T_i/dr$ times a spatial scale is an instability factor, but what that spatial scale is is not agreed upon.  It could be related to the average magnetic field-line curvature, the local shear, or the global shear in the magnetic field.  A density gradient is stabilizing, so $d \ln T/d \ln n$ should not be too large, but a sufficiently weak temperature gradient can be stable even when the density profile is flat.  

To avoid impurity accumulation, it is not clear that stabilizing the trapped electron mode when ion temperature gradient mode is unstable is beneficial.  This might make ion heat transport rapid compared to particle transport, which is bad.  What is needed is a rapid transport of non-hydrogenic ions relative to the heat transport in a plasma that primarily has hydrogenic ions.

What seems to be agreed upon is that linear instability theory is a poor surrogate for relative levels of microturbulent transport \cite{McKinney:2019}.   The ITG mode can be stabilized by zonal flows \cite{Diamond:2005}, though in stellarators wave coupling in other forms than zonal flows may be more important \cite{Hegna:2018}. 

Stellarators can be designed for optimal microturbulent transport \cite{Xanthopoulos:2014}, but such optimizations require a surrogate.  Full simulations of microturbulence are too time consuming to be practical.   The reliability of full gyrokinetic simulations is debated, but to the extent that they can be taken to be reliable they could be used (1) to test whether a given stellarator configuration has acceptable transport properties and (2) to determine which features of the magnetic configuration have the greatest effect on the microturbulent transport.  The microturbulence codes GENE, XGC, and GTC have been primarily developed for tokamaks but stellarator versions, such as XGC-S \cite{Cole:2019}, are being developed.

Plasma confinement can be greatly enhanced by transport barriers.  The best known is the H-mode enhancement of tokamak confinement by approximately a factor of two by the formation of a narrow edge pedestal  \cite{Physics Today:2019}.  The formation and the stability of this transport barrier can be strongly influenced by tokamak shaping and in particular by having negative triangularity, which has the mid-plane point of the triangle on the small major radius side of the plasma \cite{Austin:2019}.  Transport barriers can form not only at the edge but also in the body of the plasma \cite{Connor:2004}.  Internal transport barriers are associated with rational magnetic surfaces, low or  negative magnetic shear, a strong local magnetic shear, such as that produced by the Shafranov shift, and $\vec{E}\times\vec{B}$ flow shear, which has a far stronger effect on the ion than on the electron transport.  

Stellarators offer much more freedom to change the properties that control internal transport barriers than do tokamaks.   Freedom from the details of the profile of the net plasma current, including the disruptions caused by that profile, imply transport barriers are an important area for exploration.  Nevertheless, existing W7-X results, Appendix \ref{sec:exp-comparison} imply such explorations are probably not required to build an attractive stellarator reactor other than to increase the certainty that the transport in a given design is acceptable.

There should be a focus on experiments and theory that can contribute over a time scale of years, not decades, to clarify the constraints of microturbulence  on reactor design.


\section{Edge control \label{sec:edge} }

\subsection{Divertors} 

The particle exhaust from plasmas should be concentrated to the location of pumps, but this concentration makes the power loading on the walls intolerably high unless a large fraction of the power is radiated away.  

A divertor is a magnetic structure that directs the plasma particles to the locations of pumps.  A detached divertor means that radiation removes essentially all of the energy from the plasma before it contacts the wall.  

Two types of magnetic structures are being considered for divertors in stellarator reactors:  resonant and non-resonant.

\subsubsection{Resonant divertors}

A resonant divertor locates a chain of islands at the plasma edge, which requires extremely accurate control of the edge rotational transform, $\iota=1/q$, which is the twist of the magnetic field lines; $q$ is the safety factor.

W7-X has a resonant divertor \cite{7-X divertor:2019,Pedersen:2019}, so this concept is being studied as part of the W7-X program.  In particular, W7-X has demonstrated that a resonant divertor can maintain stable detachment and radiate most of the plasma energy before the plasma reaches the walls.

\subsubsection{ Non-resonant divertors} 

Non-resonant divertors use the Hamiltonian mechanics concepts of Cantori and turnstiles.  Magnetic field lines obey exactly the equations of one-and-a-half degree of freedom Hamiltonian mechanics, $H(p,q,t)$, although for field lines the three variables of the Hamiltonian mechanics are three spatial coordinates.  Beyond the outermost confining magnetic surface, a double magnetic flux tube is formed in each period of the stellarator (1/2 the flux comes in and 1/2 goes out).  The two parts of these tubes strike the wall at remarkably robust locations.  

Non-resonant divertors have been explored less than resonant divertors, but there are several theoretical papers on non-resonant divertors \cite{Stell-desgn:2015,HSX divertor:2017,CTH divertor 2018,Divertors-Boozer-Punjabi,Divertors-Punjabi-Boozer}.  Unlike resonant divertors, non-resonant divertors place no constraint on the edge rotational transform and the width of the escaping flux tube that carries plasma to the pumps can be adjusted.


\subsection{Protection of the walls from $\alpha$ particles }

Helium ions (alpha particles) produced by the nuclear reactions can become deeply embedded in the walls if they strike while still energetic.  The accumulation of helium gas in crystal lattices creates blisters and fuzzy regions, which destroys the structural integrity of the walls.

Three strategies have been proposed for addressing this issue: (1) Apply whatever constraints are necessary on the variation of the magnetic field strength on the magnetic surfaces to limit the loss of alpha particles. (2)  Design the edge magnetic field so the energetic trapped alpha particles, which are the problem, strike the wall in a location in which they harmlessly go into a liquid, such as lithium or tin, not a solid wall.  The feasibility of doing this is essentially unexplored. (3) Avoid alpha-particle damage altogether by covering plasma facing components with a thin liquid film, Section \ref{liquid}.


\section{Implications of technical developments \label{sec:technical} }

Technical developments are of particular importance in four areas (1) coils, (2) liquid films for covering first walls, (3) solid first wall materials, and (4) breeding blankets for tritium.  The design of a stellarator reactor that has open access to the plasma chamber requires a suitable choice for the coil system.  But, when this is done, a fast development of fusion requires only that an appropriate space allocation be made for the first wall, the blankets, and shields.  Several versions of these systems should be made to test various designs.  The replacement of inadequate components must be part of the research on the test reactor.

\subsubsection{Developments for coils} 

Technical developments in high-temperature superconducting coils for fusion applications was the subject of a 2018 \emph{Nuclear Fusion} review \cite{HT magnets}.  This review included a discussion of use of  joints in coils.  The construction of both tokamaks and stellarators could be faster and cheaper if coils could be delivered in pieces that are joined during device construction.   

Commonwealth Fusion Systems \cite{Whyte:2019} has placed a strong focus on developing coils for fusion systems that can operate at much higher magnetic fields than those in existing tokamaks.  This work is important for stellarators as well as tokamaks.  The required minor radius of a plasma will be found to scale as $a\propto \mathcal{D}^{2/5}/B^{4/5}$ while the total power output of a reactor $P_T$ for a given wall loading $p_w$ scales as $a^2$.  Higher magnetic fields allow power plants to be built with a smaller unit size and allow compensation for poor confinement, a large $\mathcal{D}$.

\subsubsection{Development of liquid films \label{liquid} } 

Even a thin layer of liquid on plasma-facing components can address four issues \cite{Liquid-walls:2019}.   First, the layer can eliminate the degradation of wall materials that can be produced by fusing plasma plasmas.  Examples are alpha particle degradation and the sudden flash of radiative energy that would occur if a piece of a tile fell into the plasma.  Second, flowing liquids can remove the surface heat load.  Third, somewhat thicker liquid layers can reduce the nuclear damage.  Fourth, liquid layers can reduce gradients such as temperature and stress.

\subsubsection{Development of solid walls} 

Although liquids can mitigate issues associated with plasma-facing components, solid walls are required even if there are liquids covering the walls.  Issues that must be addressed relative to materials for the first wall are discussed in \cite{Fusion materials:2017,Wall-materials:2019}

\subsubsection{Development of tritium breeding blankets} 

Major challenges and fundamentally different design choices exist for the blankets that breed the tritium burnt in fusion systems.  These are reviewed in \cite{Abdou:2015}. 

\vspace{0.2in}

\section*{Acknowledgements}

This material is based upon work supported by the U.S. Department of Energy, Office of Science, Office of Fusion Energy Sciences under Award DE-FG02-95ER54333 and by grant 601958 within the Simons Foundation collaboration  "Hidden Symmetries and Fusion Energy." 


\appendix


\section{Fusion power and transport \label{sec:fusion and transport} }

Freidberg, Mangiarotti, and  Minervini have noted \cite{Freidberg:2015} ``\emph{that the overall design of a tokamak fusion reactor is determined almost entirely by the constraints imposed by nuclear physics and fusion engineering.}"  Related constraints apply to stellarators and allow a simplified determination of the requirements of a stellarator reactor and how they depend on physics properties of the plasma, which is the subject of this appendix.   

A small unit size for fusion reactors, measured by the total power output $P_T$, is in conflict with having a high power density on the walls $p_w$ since $P_T\propto Ra p_w.$  The basic fusion concept sets the aspect ratio $R/a$, but the minor radius $a$ is determined by transport as long as the minor radius is sufficiently large compared to the thickness of the blankets and shields surrounding the plasma.  When transport would allow a minor radius smaller than this, the DT fuel mixture could be degraded from the optimal 50/50 mixture for the reactor design to be consistent with an adequate $a$.  The diffusion coefficient below which transport becomes too small is comparable to gyro-Bohm with an enhancement factor $\mathcal{D}\approx0.1$.  

The units that are used are 10~keV for temperature, $10^{20}/$m$^3$ for number density, Tesla for magnetic field, megajoules for energy, and seconds for time.  In these units, the Boltzmann coefficient, which converts $10^{20}~$particles/m$^3$ times 10~keV into mega-Jules per cubic meter, is $k_B = 10^{20} \times 1.602 \times 10^{-21} = 0.1602.$  The permeability of free space $\mu_0=4\pi\times10^{-7}$ in standard scientific units becomes $\mu_0=0.4\pi$ in the units that are used in this paper.

The radial coordinate $r$ is defined so the volume enclosed by a magnetic flux surface is $(2\pi R)(\pi r^2)$ with $R$ the major radius.  The edge of the plasma is at $r=a$, which is the standard stellarator definition of the minor radius.  The standard definition of the minor radius of a tokamak, $a_t$ has a plasma volume $\kappa_e(2\pi R)(\pi a_t^2)$, where $\kappa_e$ is the elongation.  That is $a_t=a/\sqrt{\kappa_e}$.


\subsection{Deuterium-Tritium power density}

John Wesson  \cite{Wesson:2004} gave a convenient expression for the power density of DT fusion, which holds with 10\% accuracy for temperatures between 10~keV and 20~keV,
\begin{eqnarray} 
p_{DT} &=& 0.77 n^2 T^2. \label{eq:nuclear-power}
\end{eqnarray}
with 1/5 of the energy in alpha particles and 4/5 in neutrons.  The power density in alpha particles is 
\begin{eqnarray}
p^\alpha_{DT}&=& 0.154 n^2T^2; \label{eq:alpha-power}\\
 c_{DT} &\equiv& \frac{p_{DT}^\alpha(0)}{(nT)^2}\\
&\approx&0.154.
\end{eqnarray}

The derivation of the power density, megawatts per meter cubed, begins with Equation (1.4.2) of Wesson's book \emph{Tokamaks} \cite{Wesson:2004}.  The power density in alpha particle is $p^\alpha_{DT} = n^2<\sigma v> \mathcal{E}_\alpha/4$.  The energy released in alpha particles per reaction is $\mathcal{E}_\alpha=(3.5~$MeV$)\times (1.60 \times 10^{-19}~$MJ/MeV$)=5.6\times 10^{-19}~$MJ.  The velocity weighted cross section with Maxwellian ions is approximated within 10\% accuracy for 10~keV$<T<$20~keV in Equation (1.5.4) as $<\sigma v>=1.1\times 10^{-22}T^2$ when the units of T are 10~keV.  This calculation gives $p^\alpha_{DT}$, Equation (\ref{eq:alpha-power}); a multiplication by five gives the full power density, $p_{DT}$,  Equation (\ref{eq:nuclear-power}).

The required energy confinement time to achieve ignition is
\begin{eqnarray}
\frac{3k_B nT}{\tau_E} &=& p^\alpha_{DT} \hspace{0.2in}  \mbox{   so   }   \\
nT\tau_E&=&\frac{3k_B}{c_{DT}}\approx 3.12. \label{ignition-condition}
\end{eqnarray} 
The minimum of $nT\tau_E$ is at $T$=1.4, which means at 14~keV.  What is precisely meant by $nT$ is not clear since both $n$ and $T$ depend on radius.  When central values are used in an analytic transport model, the constant 3.12 becomes 4.39, Equation (\ref{Model n tau T}).



\subsection{Transport model \label{sec:transport model} }

 The equilibrium between heat transport and fusion power in alpha particles is 
\begin{eqnarray}
\frac{1}{r} \frac{d}{dr}(r Q) &=& p^\alpha_{DT}(0) f, \mbox{   where   } \label{transport eq} \\
f&\equiv&\left( \frac{nT}{n_0T_0}\right)^2,
\end{eqnarray}
and $p^\alpha_{DT}(0)$ is the central power density provided by the fusion-produced alpha particles.  The heat flux is
\begin{eqnarray}
Q(r) &=& - 3 k_B D \frac{dnT}{dr},
\end{eqnarray}
and $D$ is the diffusion coefficient for plasma pressure. 

An integration of the transport equation across the plasma $0<r<a$ gives
\begin{equation}
aQ(a) =  p^\alpha_{DT}(0) \int_0^a f r dr.  \label{balance}
\end{equation} 

\subsubsection{Analytic model \label{sec:analytic} }

An analytic model is obtained for a diffusion coefficient that is proportional to the plasma pressure, which is the case for gyro-Bohm diffusion when the density is proportional to the square root of the temperature.  Let
\begin{eqnarray}
D(r) &=& D_0 \sqrt{f}, \hspace{0.2in} \mbox{ so  }  \\
Q(r) &=& - \frac{3}{4} p_0 D_0 \frac{df}{dr};  \label{Q(r)} \\
p_0 &\equiv& 2k_B n_0T_0,
\end{eqnarray}
where $D_0$ is a constant and $p_0$ is the central plasma pressure.

The solution to the equation for $f$ given by Equations (\ref{transport eq}) and (\ref{Q(r)}) is 
\begin{eqnarray}
f(x)&=&J_0(x),  \mbox{   with  }\\
x&\equiv&kr; \\
f(0)&=&1, \mbox{   and    } f(ka)=0.
\end{eqnarray}
 $J_0(x)$ is the zeroth order Bessel function of the first kind, which has its first zero, $J_0(\lambda_0)=0$, at $\lambda_0=2.405$...,  $dJ_0/dx=-J_1(x)$, and $d(xJ_1)/dx = x J_0(x)$.  The boundary condition $f(ka)=0$ implies
 \begin{equation}
 ka=\lambda_0.
 \end{equation}
 The energy flux at the plasma edge is
  \begin{eqnarray}
  Q(a)&=& \frac{3}{4} \frac{p_0D_0}{a} \mathcal{F}_0; \label{Q(a)} \\
  \mathcal{F}_0 &\equiv& \lambda_0 J_1(\lambda_0), \mbox{   and   } \\
 k^2 &=& \frac{ p_{DT}^\alpha(0) }{ \frac{3}{4} p_0D_0}, \mbox{  or  }  \label{k^2 exp}\\
 \frac{D_0}{ a^2} &=& \frac{ p_{DT}^\alpha(0) }{ \frac{3}{4} p_0\lambda_0^2} \\
 &=& \frac{2c_{DT}}{3k_B  \lambda_0^2}n_0T_0.   \label{D_0/a^2}
\end{eqnarray}
using Equation (\ref{eq:alpha-power}).  Equation (\ref{D_0/a^2}) gives the required confinement for the power from alpha heating to balance the thermal losses at the plasma edge.  

The thermal energy in the plasma $W_{th}$ is the integral of $3p/2$ over the volume of the plasma; 
\begin{eqnarray}
 W_{th}&=& (2\pi)^2 R \int_0^a \frac{3}{2}p_0 \sqrt{f} r dr \\
 &=& 3 p_0 \pi^2 Ra^2 \bar{m}_0 \mbox{    where   } \label{W_th} \\
\bar{m}_0 &\equiv&   \frac{ \int_0^{\lambda_0} \sqrt{J_0(x)}xdx}{\frac{1}{2}\lambda_0^2} \approx0.608. 
 \end{eqnarray}
 When performed numerically $ \int_0^{\lambda_0} \sqrt{J_0(x)}xdx\approx1.76$.  
 
 The energy confinement time is 
 \begin{eqnarray}
 \tau_E&\equiv& \frac{W_{th}}{(2\pi)^2 RaQ(a)} \\
 &=& \frac{\bar{m}_0}{\mathcal{F}_0}\frac{a^2}{D_0} \label{tau-D} \\
 &\approx& 0.488\frac{a^2}{D_0}.
 \end{eqnarray} 
 
 The requirement for alpha heating to balance the thermal losses is
 \begin{eqnarray}
 \frac{W_{th}}{\tau_E}&=& (2\pi)^2R\int p^\alpha_{DT} rdr  \mbox{   or   } \\
 \tau_E &=& \frac{\frac{3}{2}p_0\int_0^a p r dr}{p^\alpha_{DT}(0) \int_0^a p^2 r dr} \\
 &=& \frac{1}{n_0T_0} \frac{3k_B}{c_{DT}} \frac{p_0\int_0^a p r dr}{\int_0^a p^2 r dr} \\
 &=& \frac{1}{n_0T_0} \frac{3k_B}{c_{DT}} \frac{\bar{m}_0\lambda_0^2}{2\mathcal{F}},    \mbox{   so   }\\ \nonumber\\
 n_0T_0\tau_E &\approx & 3.12 \times 1.409 \approx4.39.  \label{Model n tau T}
 \end{eqnarray}


\subsubsection{Gyro-Bohm diffusion}

The implications of transport on designs of toroidal magnetic fusion systems, stellarators and tokamaks, requires a normalizing transport model.  Gyro-Bohm diffusion will be used for two reasons:  (1) An empirical scaling law $\tau _{ E}^{\rm ISS04}$, Equation (\ref{emp-scaling}), describes a broad range of stellarator and tokamak experiments, Figure \ref{fig:scaling_f}.  This scaling law is accurately approximated by gyro-Bohm scaling, Equation (\ref{tau_E gyro-Bohm}), with a dimensionless multiplying factor $\mathcal{D}$. (2) Even non-turbulent transport models, Appendix \ref{non-turbulent}, can differ by a number of orders of magnitude from gyro-Bohm transport models, either larger or smaller, but gyro-Bohm transport with $\mathcal{D}\approx1$ gives optimal reactor designs.  A heuristic derivation of the gyro-Bohm diffusion coefficient is given in Appendix \ref{gyro-Bohm derivation}.

The gyro-Bohm diffusion coefficient is
\begin{equation}
D_{gB}\equiv \rho_i^2 \frac{C_s}{a}, \label{D_gB}
\end{equation}
where $\rho_i\equiv C_s/\omega_{ci}$ is the ion gyroradius using the speed of sound $C_s\equiv\sqrt{T/m_i}$, and $a$ is the plasma minor radius.  When $D_{gB}$ is  evaluated at an ion mass of 2.5 times the proton mass,  
\begin{eqnarray}
D_{gB}&=&\frac{c_{gB}}{a} \frac{T^{3/2}}{B^2}; \\
c_{gB}&\approx& 162 \label{c_gB}
\end{eqnarray}
 The speed of sound is $C_s\equiv\sqrt{T/m_i} = 6.19\times 10^5 \sqrt{T}$ and the ion gyroradius is $\rho_i\equiv C_s/\omega_{ci} = 1.62\times 10^{-2} \sqrt{T}/B$.
 
 The analytic model of Appendix \ref{sec:analytic}, is obtained when the density profile has the form $n\propto\sqrt{T}$, then $D_{gB}/\sqrt{f}$ is constant.  
 
 To study the effect of enhanced or reduced transport, a dimensionless coefficient $\mathcal{D}$ is introduced so
 \begin{eqnarray}
 D_0 &=&\mathcal{D}D_{gB}(0), \mbox{    or    }  \\
 &=&\mathcal{D}\frac{c_{gB}}{a} \frac{T_0^{3/2}}{B^2},  \label{D_0-D_gB}
\end{eqnarray}
where the constant $c_{gB}$ is given in Equation (\ref{c_gB}).


\subsubsection{Heuristic derivation of gyro-Bohm diffusion \label{gyro-Bohm derivation} }

 The heuristic derivation of the gyro-Bohm diffusion coefficient starts with general expression for a radial diffusion coefficient,  $D\approx\Delta^2/\tau_{co}$, where $\Delta$ is the radial scale of the microturbulence, and $\tau_{co}$ is the correlation time of the flow $v_r$ that gives the radial scale; $\Delta\approx v_r\tau_{co}$.   The radial velocity in electrostatic turbulence is $v_r=\tilde{E}_\theta/B$.  The radial motion produces a change in the electric potential $\tilde\phi \approx \Delta (T/ea)$ where $T$ is the plasma temperature and the minor radius $a$ is the scale over which the temperature varies.  The poloidal variation in the potential is $\delta\Delta (T/ea)$, where $\delta\Delta$ is the poloidal variation of the radial scale.  Consequently,  $\tilde{E}_\theta\approx(\delta\Delta/\Delta_\theta)(T/ea)\approx T/ea$; the radial scale varies by roughly the poloidal scale over the poloidal scale of the turbulence, $\Delta_\theta$.  That is, $D\approx (T/eBa)\Delta$, where $T/eBa=\rho_iC_s/a$, the ion gyroradius times the speed of sound with both calculated using the temperature $T$.  Therefore, one can let
 \begin{equation}
 D=\Delta  \frac{\rho_sC_s}{a},
 \end{equation}
 where the approximations are absorbed into the radial scale size of the turbulence $\Delta$.  
 
 In gyro-Bohm diffusion, $\Delta=\rho_s$, which is a typical scale of fluctuations in ITG turbulence.  In Bohm-like diffusion, $\Delta\approx a$, which is as large as it can be.  The enhancement factor of gyro-Bohm diffusion has the interpretation
 \begin{equation}
 \mathcal{D}=\frac{\Delta}{\rho_s},
 \end{equation}
 but can also differ from unity because the plasma is not microturbulent, Appendix \ref{non-turbulent}, or turbulence is present in only part of the plasma.


\subsubsection{Gyro-Bohm scaling of $\tau_E$}

The scaling of the energy confinement time will be studied using Equation (\ref{tau-D}) with $D_0$ replaced by the gyro-Bohm-scaled diffusion coefficient, Equation (\ref{D_0-D_gB}),
\begin{equation}
\tau_E = \frac{\bar{m}_0}{c_{gB}\mathcal{F}_0} \frac{B^2a^3}{\mathcal{D} T_0^{3/2}}.
\end{equation}

\begin{figure}
\centerline{ \includegraphics[width=2.5in]{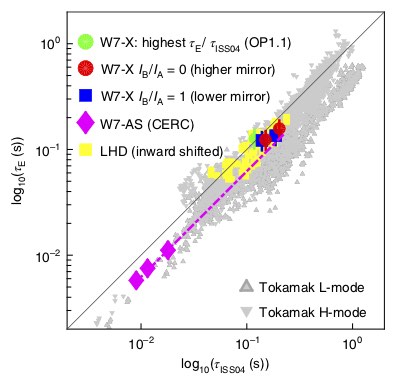} }
\caption{The energy confinement times observed in both stellarator and tokamak experiments are compared to the  stellarator scaling law, $\tau _{ E}^{\rm ISS04}$, Equation (\ref{emp-scaling}).  This figure was  Figure 4 in the 2018 Nature Physics article on W7-X  \cite{W7-X:Nature}.  }
\label{fig:scaling_f}
\end{figure}

The convention is to replace the temperature dependence of $\tau_E$ with a thermal power $P_{th}$ dependence.   Since $P_{th}=W_{th}/\tau_E$ and the central pressure is $p_0=2k_Bn_0T_0$, Equation (\ref{W_th}) implies
\begin{eqnarray}
\frac{1}{T_0} &=& 6\pi^2k_B\bar{m}_0 \frac{Ra^2}{P_{th}},   \mbox{    and  } \\
\tau_E &=& \bar{m}_0  \left(\frac{1}{\mathcal{F}_0 c_gB}\right)^{2/5} \left(6\pi^2 k_B\right)^{3/5} \nonumber\\
&& \hspace{0.2in} \times \frac{a^{12/5} R^{3/5} n_0^{3/5} B^{4/5} }{\mathcal{D}^{2/5} P_{th}^{3/5} } \\
&\approx& 0.281 \frac{a^{12/5} R^{3/5} n_0^{3/5} B^{4/5} }{\mathcal{D}^{2/5} P_{th}^{3/5} }. \label{tau_E gyro-Bohm}
\end{eqnarray}
The parameter dependencies is this formula reproduce those of the $\tau _{ E}^{\rm ISS04}$ scaling law, Equation (\ref{emp-scaling}) with surprising accuracy.  The $\tau _{ E}^{\rm ISS04}$  scaling law represents both tokamak and stellarator experiments, Figure \ref{fig:scaling_f}.

\subsubsection{Gyro-Bohm scaling of non-ignited experiments}

The magnetic field that is required to reach a central density $n_0$ and temperature $T_0$ can be calculated in terms of the thermal power, $P_{th}=(2\pi)^2Ra Q(a)$, supplied, the enhancement over gyro-Bohm diffusion, $\mathcal{D}$, and the major $R$ and minor radius, $a$.  Equation (\ref{Q(a)}) for the edge heat flux, $Q(a)$, $p_0=2k_Bn_0T_0$ for the central pressure, Equation (\ref{D_0-D_gB}) for the relation between $D_0$ and gyro-Bohm diffusion imply
\begin{eqnarray}
B &=& \sqrt{\frac{3}{2}k_B c_{gB} \mathcal{F}} \sqrt{\frac{\mathcal{D} n_0T_0^{5/2}}{a^2Q(a)} } \\
&\approx& 6.97 \sqrt{\frac{\mathcal{D} n_0T_0^{5/2}}{a^2Q(a)} } \\
&=& \sqrt{3\pi^2 k_B c_{gB} \mathcal{F}} \sqrt{\mathcal{D} n_0T_0^{5/2} \frac{R/a}{P_{th}} } \\
&\approx& 31.0  \sqrt{\mathcal{D} n_0T_0^{5/2} \frac{R/a}{P_{th}} }. \label{Non-ignited B}
\end{eqnarray}
Equivalently,
\begin{eqnarray}
\mathcal{D}&=& \frac{1}{3\pi^2k_Bc_{gB}\mathcal{F}} \frac{B^2 P_{th}}{n_0T_0^{5/2} R/a}  \\
&=&1.043\times10^{-3}  \frac{B^2 P_{th}}{n_0T_0^{5/2} R/a}.
\end{eqnarray}


\subsubsection{Gyro-Bohm scaling of ignited experiments}

When the plasma is undergoing a steady fusion burn, Equation (\ref{D_0/a^2}) gives an expression for the required $D_0/a^2$.  The expression obtained for $D/a^2$ from gyro-Bohm scaling is given by Equation (\ref{D_0-D_gB}).  Equating these two expressions provides an expression for the central density, $n_0=n_b$, with
\begin{eqnarray}
n_b&=&\frac{3k_B c_{gB} \lambda_0^2}{2c_{DT}}\frac{\mathcal{D}\sqrt{T_0}}{B^2a^3}\\ \label{n_b}
&=&252.8\lambda_0^2\frac{\mathcal{D}\sqrt{T_0}}{B^2a^3}.
\end{eqnarray}
One less parameter is required to describe ignited than non-ignited experiments.  

Equation (\ref{Q(a)}) for the edge heat flux, $Q(a)$, $p_0=2k_Bn_0T_0$ for the central pressure, Equation (\ref{D_0-D_gB}) for the relation between $D_0$ and gyro-Bohm diffusion, and $n_0=n_b$ using Equation (\ref{n_b}) imply that in an ignited plasma
\begin{eqnarray}
B&=&\left(\frac{9}{4} \frac{c_{gB}^2k_B^2}{c_{DT}} \lambda_0^2\mathcal{F}_0\right)^{1/4} \left(\mathcal{D}^2 \frac{T_0^3}{a^5 Q(a)}\right)^{1/4} \\
&=& 9.96\left(  \lambda_0^2\mathcal{F}_0  \right)^{1/4} \left( \frac{\mathcal{D}^2 T_0^3}{a^5 Q(a)}\right)^{1/4}.
\end{eqnarray}


\subsection{Transport with a confining annulus \label{sec:annulus} }

\begin{table}
  \centering 
  \begin{eqnarray}
 \left(\begin{array}{cccccc}\alpha & 0 & 0.5 & 0.6 & 0.7 & 0.8 \\ \lambda(\alpha) & 2.405 & 2.554 & 2.703 & 2.963 & 3.455\\ \bar{m}(\alpha) & 0.6 & 0.7 & 0.8 & 0.8 & 0.9 \\ \mathcal{F}(\alpha) & 1.248 & 1.930 & 2.400 & 3.211 & 4.859\end{array}\right)
  \nonumber \end{eqnarray}
  \caption{ The argument $\lambda$ of the Bessel functions at the plasma edge, $r=a$, the ratio $\bar{m}$ of the average to the central pressure, and  the enhancement of the fusion power $\mathcal{F}$ are given as a function of $\alpha$, which is the fraction of the plasma radius in which diffusion is assumed to go to infinity.  }\label{Values-lambda-F}
\end{table}

A confining annulus means that the diffusion coefficient $D(r)$ is extremely large in the central part of the plasma $0<r<\alpha a$, so $f=1$ there, but within the confining annulus $\alpha a <r <a$, the diffusion coefficient has the same form as in Appendix \ref{sec:transport model}, $D(r) = D_0 \sqrt{f}$.  The solution for $f$ in the confining annulus $\alpha a <r <a$ is
\begin{eqnarray}
f(x) &=&\frac{Y_0(\lambda) J_0(x)-J_0(\lambda)Y_0(x)}{J_0(\alpha\lambda)Y_0(\lambda) - J_0(\lambda)Y_0(\alpha\lambda)},  \label{alpha soln}
\end{eqnarray}
where $k^2=4p^\alpha_{DT}(0)/(3p_0D_0)$ as before, Equation (\ref{k^2 exp}).  The boundary conditions are $f(\alpha\lambda)=1$ and $f(\lambda)=0$.  $J_0(x)$ and $K_0(x)$ are the Bessel functions of the first and second kind.  Both obey the relations $dJ_0/dx=-J_1(x)$ and $d(xJ_1)/dx = x J_0(x)$.

The function $\lambda(\alpha)$ is given implicitly by  Equation (\ref{balance}), which is obtained by equating the total exiting heat flux $2\pi a Q(a)$ per unit length of the plasma in the toroidal direction with the total alpha-heating power per unit length.
\begin{eqnarray}
 \int_0^\lambda f xdx &=& \frac{(\alpha \lambda)^2}{2} +\mathcal{F}(\lambda) \nonumber\\
&& - \alpha\lambda \frac{Y_0(\lambda) J_1(\alpha\lambda)-J_0(\lambda)Y_1(\alpha\lambda)}{J_0(\alpha\lambda)Y_0(\lambda) - J_0(\lambda)Y_0(\alpha\lambda)}; \hspace{0.3in}\\
\mathcal{F}(\alpha)&\equiv&  \lambda  \frac{J_1(\lambda)Y_0(\lambda) - J_0(\lambda) Y_1(\lambda)}{J_0(\alpha\lambda)Y_0(\lambda) - J_0(\lambda)Y_0(\alpha\lambda)};\\
ka\left(\frac{df}{dx}\right)_\lambda &=& - \mathcal{F}(\alpha).
\end{eqnarray}
The implication is that Equation (\ref{Q(a)}) for $Q(a)$ holds when $\mathcal{F}_0$ is replaced by $\mathcal{F}(\alpha)$.  Indeed, $\mathcal{F}$ can be defined by Equation (\ref{Q(a)}).


\begin{table}
  \centering 
  \begin{eqnarray}
\left(\begin{array}{ccccccc} & \alpha = & 0 & \alpha= & 0.5 & \alpha = & 0.7 \\\frac{r}{a} & \frac{p}{p_0} & \frac{d\ln p}{d\ln r} & \frac{p}{p_0} & \frac{d\ln p}{d\ln r} & \frac{p}{p_0} & \frac{d\ln p}{d\ln r} \\0.5 & 0.818 &0.448 & 1 & 0.408 & 1 & 0 \\0.6 & 0.737 & 0.729 & 0.907 & 0.695 & 1 & 0 \\0.7 & 0.638 & 1.191 & 0.789 & 1.163 & 1 & 1.075 \\0.8 & 0.518 &2.084 & 0.642 &2.063 & 0.822 & 2.000 \\0.85 & 0.445 & 2.952 & 0.553 & 2.936 & 0.7105 & 2.887 \\0.9 & 0.361 & 4.66 & 0.448 & 4.647 & 0.577& 4.613\end{array}\right)
  \nonumber \end{eqnarray}
  \caption{ The radial profiles of the pressure and the logarithmic derivative of the pressure with respect to radius are gives for three values $\alpha$, which is the fraction of the plasma radius in which diffusion is assumed to go to infinity.  As $\alpha$ becomes larger, the stability measure $d\ln p/d\ln r$ becomes smaller at a given radius, but the pressure at which $d\ln p/d\ln r$ reaches a certain value becomes larger.   }\label{Profiles}
\end{table}

Equation (\ref{balance}) for energy balance is satisfied when
\begin{eqnarray}
&&0 = \frac{\alpha \lambda}{2}  +  \frac{Y_0(\lambda) J_1(\alpha\lambda)-J_0(\lambda)Y_1(\alpha\lambda)}{J_0(\alpha\lambda)Y_0(\lambda) - J_0(\lambda)Y_0(\alpha\lambda)}; \label{alpha-lambda} \hspace{0.3in}\\
&& aQ(a)= p^\alpha_{DT} \frac{\lambda}{k^2}\mathcal{F}(\alpha).
\end{eqnarray}
Equatiion (\ref{alpha-lambda}) implicitly gives the function $\lambda(\alpha)$, Table \ref{Values-lambda-F}.  In the absence of a region of rapid transport, $\alpha=0$, the solution vanishes, $f(\lambda)=0$ at $\lambda=\lambda_0\approx2.405$ and $\mathcal{F}=\lambda_0J_1(\lambda_0)$.

The equation for energy balance $\int_0^a p_{DT}^\alpha rdr=aQ(a)$ can be used to define $\lambda$ for any pressure profile that satisfies the transport equation as 
\begin{equation}
\lambda^2 \equiv \frac{p_{DT}^\alpha(0) a^2 \mathcal{F}}{\int_0^a p_{DT}^\alpha(r) r dr}.
\end{equation}
Similarly $\bar{m}$ can be defined as $\bar{m}\equiv 2\int_0^a p(r)rdr/p_0a^2$.  These definitions give the equations derived in Appendix \ref{sec:transport model} general validity.

Plasmas are generally unstable to microturbulence when the logarithmic gradient of the pressure becomes large compared to unity;
\begin{eqnarray}
-\frac{d \ln(p)}{\ln(r)} & = & \frac{x}{2} \frac{df/dx}{f} \\
&=&  \frac{x}{2} \frac{Y_0(\lambda) J_1(x)-J_0(\lambda)Y_1(x)}{J_0(\alpha\lambda)Y_0(\lambda) - J_0(\lambda)Y_0(\alpha\lambda)}. \hspace{0.3in}
\end{eqnarray}
The pressure profile and the profile of the logarithmic derivative of the pressure are given in Table \ref{Profiles}.


\subsection{Comparison with experiments \label{sec:exp-comparison} }

The observed global energy confinement in stellarator experiments is summarized by the scaling \cite{Yamada:2005},
\begin{equation}
\tau _{ E}^{\rm ISS04} = 0.134\frac{a^{2.28}R^{0.64}}{P^{  0.61}} \bar{n}_{e}^{0.54} B^{0.84}\iota_{2 / 3}^{0.41}, \label{emp-scaling}
\end{equation}
where the energy confinement time is in seconds, the minor $a$ and the major radius $R$  are in meters, the volume averaged magnetic field $B$ is in Tesla, the volume-averaged electron density $\bar{n}$ is in $10^{20}/$m$^3$, the effective heating power $P$ is in mega-Watts, and the rotational transform $\iota_{2/3}$ is at a radius $r=2a/3$.  The minor radius $a$ is defined so the plasma volume is $(2\pi R)(\pi a^2)$.  

The $\tau _{ E}^{\rm ISS04}$ scaling law represents both tokamak and stellarator experiments, Figure \ref{fig:scaling_f}, though tokamak H-mode experiments have up to a factor of two better confinement than predicted.  Paradoxically, the radial dependence of the transport seen in W7-X, as reported in the article from which this figure was taken \cite{W7-X:Nature}, does not agree with that expected for gyro-Bohm transport.  Nevertheless, the overall dependencies of the $\tau _{ E}^{\rm ISS04}$ scaling law are given by gyro-Bohm transport, Equation (\ref{tau_E gyro-Bohm}).   The coefficient in the stellarator scaling is a factor of 2.09 times smaller than in gyro-Bohm scaling, which can be counterbalanced by $\mathcal{D}=6.3$.  The rotational-transform dependence of $\tau _{ E}^{\rm ISS04}$ can be interpreted as $\mathcal{D}=\Delta/\rho_s \propto 1/\iota$, which may even be correct.

The detached divertor experiments in the Large Helical Device (LHD) that were reported in 2018 \cite{LHD:2108} had $a=0.55$, $R=3.90$, $n_0=0.7$, $B=3$, $P_{th}=9$, and a stored plasma energy $W_{th}=0.35$.  These results were said to be agreement with Equation (\ref{emp-scaling}) for stellarator scaling.  The central temperature is related to the thermal energy content in the analytic model by
\begin{eqnarray}
T_0 &=& \frac{W_{th}}{3\pi^2\bar{m}_0 k_B n_0Ra^2} \\
&\approx& 0.147.
\end{eqnarray} 
Consistency with Equation (\ref{Non-ignited B}) is obtained for $\mathcal{D}=2.1$.  

If $\mathcal{D}$ were 2.1 for stellarators, but the energy confinement time were factor of two longer, as in the case in H-mode tokamaks in Figure \ref{fig:scaling_f}, then $\mathcal{D}$ would be 0.37 for H-mode tokamaks.  Smaller values of $\mathcal{D}$ have been seen in tokamak and stellarator experiments.

A study of long-pulse DIII-D results published in 2018 \cite{DIII-D:2018} had $T_0=(T_e+T_i)/2=0.45$ and $n_0=0.5$, $B=1.6$, $R=1.7$,  $a_t=0.6$, and $P_{th}=15.6$.  The elongation was $\kappa_e =2$, which makes the stellarator definition $a=\sqrt{\kappa_e}a_t=0.849$.  A fit gives $\mathcal{D}=0.31$.

Early results from W7-X \cite{7-X divertor:2019,Pedersen:2019} demonstrate that excellent confinement can be obtained, $\mathcal{D}=0.05$, though this confinement rapidly degrades, possibly because continual pellet injection is not yet available.  The central plasma has $T_i=T_e=3.5~$keV and $n_0=0.8\times10^{20}/\mbox{m}^3$, $B=2.5~$T, $R=5.5~$m,  $a=0.5~$m, and $P_{th}=5~$MW.  W7-X was able to maintain plasma parameters for ten seconds \cite{Klinger:2019} with $T_i=T_e=1.9~$keV and $n_0=1.6\times10^{20}/\mbox{m}^3$, $B=2.5~$T, $R=5.5~$m,  $a=0.5~$m, and $P_{th}=5.9~$MW. These results give $\mathcal{D}=0.13$, a value that yields attractive reactor designs.


Burning plasma experiments in ITER seem to require a value of $\mathcal{D}$ consistent with those seen in DIII-D.  For example, the burning-plasma scenario outlined in Table 1 of \cite{Green:2003} for ITER had $P_T=500$, $a_t=2$, $R=6.2$, $\kappa_e=1.8$, $B=5.3$, $<n>=1.1$, and $<T>=0.89$.  Assuming broad profiles so $n_0=1.18$ and $T_0=1.1$ gives $\mathcal{D}=0.42$.  Similarly, the European Union design for a pulsed demonstration (DEMO) tokamak reactor \cite{EU-DEMO2019} has  $T_0=~25$keV and $n_0=1.5\times10^{20}/\mbox{m}^3$, $B=5.9~$T, $R=9~$m,  $a_t=2.9~$m, and $P_T=2,014~$MW, assuming central values are twice their volume averages.  The stellarator equivalent minor radius is $a=\sqrt{\kappa_e}a_t=3.67$~m. This requires $\mathcal{D}=0.11$.  The power loading is $p_w=P_T/((2\pi)^2 \kappa_e R a)=1.2~$MW/m$^2$, where $\kappa_e=1.6$ is the elongation.

When the confinement factor $\mathcal{D}$, the magnetic field strength $B$, and the wall loading $p_w$ are held constant, the factor that determines the total power output $P_T$ scales as $a^2\propto T_0^{6/5}$.  The higher plasma temperature required even in a pulsed tokamak reactor offsets its lower aspect ratio in comparison to a steady state stellarator reactor.  For example using stellarator definitions, the European Union DEMO has an aspect ratio of 2.45 but a central temperature is 2.5 times greater than would probably be chosen for a stellarator reactor, $2.45 \times(2.5)^{6/5}=7.364$, which is a reasonable aspect ratio for a stellarator reactor.

\vspace{0.1in}


\subsection{Non-turbulent transport \label{non-turbulent} }

The characteristic diffusion coefficient for neoclassical transport in which the particle drift trajectories make small excursions from the magnetic surfaces is 
\begin{eqnarray}
D_{nc} &=& \alpha_{nc} \rho_i^2\nu_i, \mbox{   and  } \\
\mathcal{D}_{nc} &\equiv& \frac{D_{nc}}{D_{gb}} \\
&=& \alpha_{nc} \frac{a}{\lambda_i},
\end{eqnarray} 
where $\alpha_{nc}$ is a dimensionless coefficient, which can be of order $10^2$ and $\nu_i$ is the ion collision frequency.  

The mean free path $\lambda_i \equiv C_s/\nu_i \approx 10.05 \times 10^3 T^2/n,$  so $\lambda_i/a \sim 5\times 10^3$ in a fusion reactor

Non-turbulent transport scales differently when the drift of some of the particles away from the magnetic surfaces is limited only by collisions.  The characteristic transport coefficient for this type of transport is
\begin{eqnarray}
D_{1/\nu} &=& \alpha_{1/\nu}\left(\frac{\rho_i}{a}\frac{C_s}{\nu_i}\right)^2\nu_i, \mbox{   and  } \\
\mathcal{D}_{1/\nu} &\equiv& \frac{D_{1/\nu}}{D_{gb}} \\
&=& \alpha_{1/\nu} \frac{\lambda_i}{a}.
\end{eqnarray} 
That is $D_{1/\nu}$ can be orders of magnitude greater than $D_{gB}$.




\begin{thebibliography}{99}

\bibitem{wind-solar fraction} \emph{The Costs of Decarbonisation: System Costs with High Shares of Nuclear and Renewables}, OECD 2019 NEA No. 7299, (Nuclear Energy Agency, Organization for Economic Co-operation and Development), \url{http://www.oecd-nea.org/ndd/pubs/2019/7299-system-costs.pdf}

\bibitem{NAS-CO2} National Academies of Sciences, Engineering and Medicine, \emph{Negative Emissions Technologies and Reliable Sequestration: A Research Agenda} National Academies Press, Washington, D.C. (2019). \url{https://www.nap.edu/read/25259/chapter/1}

\bibitem{SCC} P. Wang, X. Deng, H. Zhou, and S. Yu, \emph{Estimates of the social cost of carbon: A review based on meta-analysis}, Journal of Cleaner Production \textbf{209}, 1494 (2019).

\bibitem{UN:energy} \emph{2019 Energy Statistics Pocketbook}, United Nations Department of Economic and Social Affairs Statistics Division (United Nations New York, 2019), available online at \url{https://doi.org/10.18356/1df8f86d-en}.

\bibitem{Fasihi:2019} M. Fasihi, O. Efimova, C. Breyer, \emph{Techno-economic assessment of CO$_2$ direct air capture plants}, Journal of Cleaner Production \textbf{224}, 957 (2019).

\bibitem{IMF}  \emph{2019 World Economic Outlook: Growth Slowdown, Precarious Recovery} (International Monetary Fund, Washington, DC, April 2018).  Available on line at ISSN 1564-5215.

\bibitem{grid-stability} C. Seneviratne and C. Ozansoy, \emph{Frequency response due to a large generator loss with the increasing penetration of wind/PV generation---A literature review}, Renewable and Sustainable Energy Reviews \textbf{57} (2016) 659 (2016).

\bibitem{overnight} U.S. Energy Information Administration, \emph{Assumptions to the Annual Energy Outlook 2020: Electricity Market Module}, \url{https://www.eia.gov/outlooks/aeo/assumptions/pdf/electricity.pdf}, January 2020.

\bibitem{NSCC:2018} D. A. Gates, D. Anderson, S. Anderson, M. Zarnstorff, D. A. Spong, H. Weitzner, G. H. Neilson, D. Ruzic, D. Andruczyk, J. H. Harris, H. Mynick, C. C. Hegna, O. Schmitz, J. N. Talmadge, D. Curreli, D. Maurer, A. H. Boozer, S. Knowlton, J. P. Allain, D. Ennis, G. Wurden, A. Reiman, J. D. Lore, M. Landreman, J. P. Freidberg, S. R. Hudson, M. Porkolab, D. Demers, J. Terry, E. Edlund, S. A. Lazerson, N. Pablant, R. Fonck, F. Volpe, J. Canik, R. Granetz, A. Ware, J. D. Hanson, S. Kumar, C. Deng, K. Likin, A. Cerfon, A. Ram, A. Hassam, S. Prager, C. Paz-Soldan,  M. J. Pueschel, I. Joseph, A. H. Glasser, \emph{Stellarator Research Opportunities: A Report of the National Stellarator Coordinating Committee}, Journal of Fusion Energy \textbf{37}, 51 (2018).


\bibitem{Physics Today:2019} R. Hawryluk and H. Zohm, \emph{The challenge and promise of studying burning plasmas: Answers to open questions that will be addressed by the ITER experiment should enable the production of fusion energy}, Physics Today \textbf{72}, issue 12, page 34 (December 2019).
 
\bibitem{EU-DEMO2019} G. Federici, C. Bachmann, L. Barucca, C. Baylard, W. Biel, L.V. Boccaccini, C. Bustreo, S. Ciattaglia, F. Cismondi, V. Corato, C. Day, E. Diegele, T. Franke, E. Gaio, C. Gliss, T. Haertl, A. Ibarra, J. Holden, G. Keech1, R. Kembleton, A. Loving, F. Maviglia, J. Morris, B. Meszaros, I. Moscato, G. Pintsuk, M. Siccinio, N. Taylor, M. Q. Tran, C. Vorpahl, H. Walden, and J.H. You, \emph{Overview of the DEMO staged design approach in Europe}, Nucl. Fusion \textbf{59}, 066013 (2019). 

\bibitem{ITER physics} F. Perkins, D. E. Post, N. A. Uckan, M. Azumi, D. J. Campbell, N. Ivanov, N. R. Sauthoff, M. Wakatani, W. M. Nevins, M. Shimada, J. Van Dam, D. Boucher, G. Cordey, A. Costley, J. Jacquinot, G. Janeschitz, S. Mirnov, V. Mukhovatov, G. Porter, S. Putvinski, M. Shimada, R. Stambaugh, M. Wakatani, J. Wesley, K. Young, R. Aymar, Y. Shimomura, D. Boucher,N. Fujisawa, Y. Igitkhanov, A. Kukushkin, V. Mukhovatov, S. Putvinski, M. Rosenbluth, and J. Wesley, \emph{Chapter 1: Overview and summary}, Nucl. Fusion \textbf{39}, 2137 (1999). 

\bibitem{Grieger:1992} G. Grieger, W. Lotz, P. Merkel, J. N\"uhrenberg, J. Sapper, E. Strumbeger, H. Wobig, R. Burhenn, V. Erckmann, U. Gasparino, L. Giannoe, H. J. Hartfuss, R. Jaenicke, G. Kuhner, H. Ringler, A. Weller, and F. Wagner, \emph{Physics optimization of stellarators}, Phys. Fluids B \textbf{4}, 2081 (1992).

\bibitem{Hender:2007} T.C. Hender, J.C Wesley, J. Bialek, A. Bondeson, A.H. Boozer, R.J. Buttery, A. Garofalo, T.P Goodman, R.S. Granetz, Y. Gribov, O. Gruber, M. Gryaznevich, G. Giruzzi, S. Gunter, N. Hayashi, P. Helander, C.C. Hegna, D.F. Howell, D.A. Humphreys, G.T.A. Huysmans, A.W. Hyatt, A. Isayama, S.C. Jardin, Y. Kawano, A. Kellman, C. Kessel, H.R. Koslowski, R.J. La Haye, E. Lazzaro, Y.Q. Liu, V. Lukash, J. Manickam, S. Medvedev, V. Mertens, S.V. Mirnov, Y. Nakamura, G. Navratil, M. Okabayashi, T. Ozeki, R. Paccagnella, G. Pautasso, F. Porcelli, V.D. Pustovitov, V. Riccardo, M. Sato, O. Sauter, M.J. Schaffer, M. Shimada, P. Sonato, E.J. Strait, M. Sugihara, M. Takechi, A.D. Turnbull, E. Westerhof, D.G. Whyte, R. Yoshino, H. Zohm and the ITPA MHD, Disruption and Magnetic Control Topical Group, \emph{Chapter 3: MHD stability, operational limits and disruptions}, Nucl. Fusion \textbf{47}  S128 (2007). 

\bibitem{Hesslow:2019} L. Hesslow, O. Embreus, O. Vallhagen, and T. F\"ul\"op, \emph{Influence of massive material injection on avalanche runaway generation during tokamak disruptions}, Nucl. Fusion \textbf{59}, 084004 (2019).

\bibitem{Breizman:2019} B. N. Breizman, P. Aleynikov, E. M. Hollmann, and M. Lehnen, \emph{Review: Physics of runaway electrons in tokamaks}, Nucl. Fusion \textbf{59} 083001  (2019).

\bibitem{tungsten flakes:2019} M. L. Reinke, S. Scott, R. Granetz, J.W. Hughes, S.G. Baek, S. Shiraiwa, R. A. Tinguely, S. Wukitch, and The Alcator C-Mod Team, \emph{Avoidance of impurity-induced current quench using lower hybrid current drive}, Nucl. Fusion \textbf{59}, 066003 (2019). 

\bibitem{Fusion materials:2017} C. Linsmeier, M. Rieth, J. Aktaa, T. Chikada, A. Hoffmann, J.  Hoffmann, A.  Houben, H. Kurishita, X. Jin, M. Li, A. Litnovsky, S. Matsuo, A. von Muller, V.  Nikolic, T. Palacios, R. Pippan, D. Qu, J. Reiser, J. Riesch, T. Shikama, R. Stieglitz, T. Weber, S. Wurster, J.-H You, and Z. Zhou, \emph{Development of advanced high heat flux and plasma-facing materials}, Nuclear Fusion \textbf{57}, 092007 (2017). 

\bibitem{Wall-materials:2019} J.W. Coenena, Y. Maoa, S. Sistlac, A. V. M\"ullerb, G. Pintsuka, M. Wirtza, J. Rieschb, T. Hoeschenb, A. Terraa, J.-H. Youb, H. Greunerb, A. Kretera, Ch. Broeckmannc, R. Neub, Ch. Linsmeiera, \emph{Materials development for new high heat-flux component mock-ups for DEMO}, Fusion Engineering and Design \textbf{146}, 1431 (2019).

\bibitem{Liquid-walls:2019} C. E. Kessel, D. Andruczyk, J. P. Blanchard, T. Bohm, A. Davis, K. Hollis, P. W. Humrickhouse, M. Hvasta, M. Jaworski, J. Jun, Y. Katoh, A. Khodak, J. Klein, E. Kolemen, G. Larsen, R. Majeski, B. J. Merrill, N. B. Morley, G. H. Neilson, B. Pint, M. E. Rensink, T. D. Rognlien, A. F. Rowcliffe, S. Smolentsev, M. S. Tillack, L. M. Waganer, G. M. Wallace, P. Wilson, and S.-J. Yoon, \emph{Critical Exploration of Liquid Metal Plasma-Facing Components in a Fusion Nuclear Science Facility}, Fusion Science and Technology \textbf{75}, 886 (2019).

\bibitem{Abdou:2015} M. Abdou, N. B. Morley, S. Smolentsev, A. Ying, S. Malang, A Rowcliffe, and M. Ulrickson, \emph{Blanket/first wall challenges and required R\&D on the pathway to DEMO}, Fusion Engineering and Design \textbf{100}, 2 (2015).


\bibitem{Stell-desgn:2015}  A. H. Boozer, \emph{Stellarator design}, J. Plasma Phys., \textbf{81}, 515810606 (2015). 

\bibitem{demountable coils} F. J. Mangiarotti and J. V. Minervini, \emph{Advances on the Design of Demountable Toroidal Field Coils With REBCO Superconductors for an ARIES-I Class Fusion Reactor},  IEEE Transaction on Applied Superconductivity \textbf{25},  4201905 (2015).

\bibitem{Yamaguchi:2019}  H. Yamaguchi, \emph{A quasi-isodynamic magnetic field generated by helical coils},  Nucl. Fusion \textbf{59} 104002 (2019).

\bibitem{Landreman:2016} M. Landreman and A. H. Boozer, \emph{Efficient magnetic fields for supporting toroidal plasmas}, Phys. Plasmas \textbf{23}, 032506 (2016).

\bibitem{Boozer:RMP} A. H. Boozer, \emph{Physics of magnetically confined plasmas}, \textbf{76}, 1071 (2004).

\bibitem{Mercier:1964} C. Mercier, \emph{Equilibrium and stability of a toroidal magnetohydrodynamic system in the neighbourhood of a magnetic axis}, Nucl. Fusion \textbf{4}, 213 (1964).

\bibitem{Garren:1991} D. A. Garren and A. H. Boozer, \emph{Existence of quasihelically symmetrical stellarators}, Phys. Fluids B \textbf{3}, 2822 (1991).

\bibitem{Landreman:2019} M. Landreman, W. Sengupta, and G. G. Plunk, \emph{Direct construction of optimized stellarator shapes. Part 2. Numerical quasisymmetric solutions}, J. Plasma Phys. \textbf{85}, 905850103  (2019).

\bibitem{Boozer:1983} A. H. Boozer, \emph{Transport and isomorphic equilibria}, \textbf{26}, 496 (1983).

\bibitem{Cary:1997} J. R. Cary and S. G. Shasharina \emph{Omnigenity and quasihelicity in helical plasma confinement systems}, Phys. Plasmas \textbf{4}, 3323 (1997).

\bibitem{Landreman:2012} M. Landreman and P.  J. Catto, \emph{Omnigenity as generalized quasisymmetry}, Phys. Plasmas \textbf{19}, 056103 (2012).


\bibitem{stell-surf-Boozer2019}  A. H. Boozer, \emph{Curl-free magnetic fields for stellarator optimization}, Phys. Plasmas, \textbf{26}, 102504 (2019)

\bibitem{Boozer:Nature2018} A. H. Boozer, \emph{Enhanced control}, Nature Physics \textbf{14}, 1157 (2018).

\bibitem{Park:2018} J. K. Park, Y. Jeon, Y. In, J. W. Ahn, R. Nazikian, G. Park, J. Kim, H. Lee, W. Ko, H. S. Kim, N. C. Logan, Z. R. Wang, E. A. Feibush, J. E. Menard, and M. C. Zarnstorff, \emph{3D field phase-space control in tokamak plasmas}, Nature Physics \textbf{14}, 1223 (2018).

\bibitem{Helander:2015} P. Helander, T. Bird, F. Jenko, R. Kleiber, G. G. Plunk, J. H. E. Proll, J Riemann, and P. Xanthopoulos, \emph{Advances in stellarator gyrokinetics}, Nucl. Fusion \textbf{55}, 053030 (2015). 

\bibitem{McKinney:2019} I. J. McKinney, M. J. Pueschel, B. J. Faber, C. C. Hegna,  J. N. Talmadge, D. T. Anderson, H. E. Mynick, and P. Xanthopoulos,  \emph{A comparison of turbulent transport in a quasi-helical and a quasi-axisymmetric stellarator}, J. Plasma Phys. \textbf{85}, 905850503 (2019).

\bibitem{Diamond:2005} P. H. Diamond, S. I. Itoh, K. Itoh, and T. S. Hahm, \emph{Zonal flows in plasma - a review}, Plasma Phys. Control Fusion, \textbf{47}, R35 (2005).

\bibitem{Hegna:2018} C. C. Hegna, P. W. Terry, and B. J. Faber, \emph{Theory of ITG turbulent saturation in stellarators: Identifying mechanisms to reduce turbulent transport}, Phys. Plasmas \textbf{25} 022511 (2018).

\bibitem{Xanthopoulos:2014} P. Xanthopoulos, H. E. Mynick, P. Helander, Y. Turkin, G. G. Plunk, F. Jenko, T. Goerler, D. Told, T. Bird, and J. H. E. Proll, \emph{Controlling Turbulence in Present and Future Stellarators}, Phys. Rev. Lett. \textbf{113}, 155001 (2014).


\bibitem{Cole:2019} M. D. J. Cole, R. Hager, T. Moritaka, J. Dominski, JR. Kleiber, S. Ku, S. Lazerson, J. Riemann, and C. S. Chang, \emph{Verification of the global gyrokinetic stellarator code XGC-S for linear ion temperature gradient driven modes}, Phys. Plasmas \textbf{26}, 082501 (2019).

\bibitem{Austin:2019} M. E. Austin, A. Marinoni, M. L. Walker, M. W. Brookman, J. S. deGrassie, A. W. Hyatt, G. R. McKee,
C. C. Petty, T. L. Rhodes, S. P. Smith, C. Sung, K. E. Thome, and A. D. Turnbull, \emph{Achievement of Reactor-Relevant Performance in Negative Triangularity Shape in the DIII-D Tokamak}, Phys. Rev. Lett. \textbf{122}, 115001 (2019).

\bibitem{Connor:2004} J. W. Connor, T. Fukuda, X. Garbet, C. Gormezano, V. Mukhovatov, M. Wakatani, the ITB Database Group and the ITPA Topical Group on Transport and Internal Barrier Physics, \emph{A review of internal transport barrier physics for steady-state operation of tokamaks}, Nucl. Fusion \textbf{44}, R1 (2004).

\bibitem{7-X divertor:2019}  T. S. Pedersen, R. Konig, M. Jakubowski, M. Krychowiak, D. Gradic,  C. Killer, H. Niemann, T. Szepesi, U. Wenzel, A. Ali, G. Anda, J. Baldzuhn,  C. Biedermann, B.D. Blackwell, H.-S. Bosch, S. Bozhenkov, R. Brakel, S. Brezinsek, J. Cai, B. Cannas, J.W. Coenen, J. Cosfeld, A. Dinklage, T. Dittmar, P. Drewelow, P. Drews, D. Dunai, F. Effenberg, M. Endler, Y. Feng, J. Fellinger, O. Ford, H. Frerichs, G. Fuchert, Y. Gao, J. Geiger, A. Goriaev, K. Hammond, J. Harris, D. Hathiramani, M. Henkel, Ye. O. Kazakov, A. Kirschner, A. Knieps, M. Kobayashi, G. Kocsis, P. Kornejew, T. Kremeyer, S. Lazerzon, A. LeViness, C. Li, Y. Li, Y. Liang, S. Liu, J. Lore, S. Masuzaki, V. Moncada, O. Neubauer, T. T. Ngo, J. Oelmann, M. Otte, V. Perseo, F. Pisano, A. Puig Sitjes, M. Rack, M. Rasinski, J. Romazanov, L. Rudischhauser, G. Schlisio, J.C. Schmitt, O. Schmitz, B. Schweer, S. Sereda, M. Sleczka, Y. Suzuki, M. Vecsei, E. Wang, T. Wauters, S. Wiesen, V. Winters, G.A. Wurden, D. Zhang, S. Zoletnik and the W7-X Team,  \emph{First divertor physics studies in Wendelstein 7-X}, Nucl. Fusion \textbf{59}, 096014 (2019).

 \bibitem{Pedersen:2019} T. S. Pedersen, R. K\"onig, M. Krychowiak, M. Jakubowski, J. Baldzuhn, S. Bozhenkov, G. Fuchert, A. Langenberg, H. Niemann, D. Zhang, K. Rahbarnia, H.-S. Bosch, Y. Kazakov, S. Brezinsek, Y. Gao, N. Pablant, and the W7-X Team, \emph{First results from divertor operation in Wendelstein 7-X}, Plasma Phys. Control. Fusion \textbf{61} 014035 (2019).


\bibitem{HSX divertor:2017} A. Bader, A. H. Boozer, C. C. Hegna, S. A. Lazerson, and J. C. Schmitt, \emph{HSX as an example of a resilient non-resonant divertor}, Phys. Plasmas \textbf{24}, 032506 (2017).

\bibitem{CTH divertor 2018} A. Bader, C. C. Hegna, M. Cianciosa, and G. J. Hartwell, \emph{Minimum magnetic curvature for resilient divertors using Compact Toroidal Hybrid geometry}, Plasma Phys. Control. Fusion \textbf{60}, 054003 (2018).

\bibitem{Divertors-Boozer-Punjabi} A. H. Boozer and A. Punjabi, \emph{Simulation of stellarator divertors}, Phys. Plasmas \textbf{25 }, 092505 (2018).

\bibitem{Divertors-Punjabi-Boozer} A. Punjabi and A. H. Boozer, \emph{Simulation of non-resonant stellarator divertor}, Phys. Plasmas \textbf{27}, 012503 (2020).

\bibitem{HT magnets} P. Bruzzone, W. H. Fietz, J. V. Minervini, M. Novikov, N. Yanagi, Y. Zhai, and J. Zheng, \emph{Review: High temperature superconductors for fusion magnets}, Nucl. Fusion \textbf{58}  103001 (2018).

\bibitem{Whyte:2019} D. Whyte, \emph{Small, modular and economically attractive fusion enabled by high temperature superconductors}, Philosophical transactions of the Royal Society of London. Series A: Mathematical, physical, and engineering sciences \textbf{377}, issue 2141, page 20180354 (2019).


\bibitem{Freidberg:2015} J. P. Freidberg, F. J. Mangiarotti, and J. Minervini, \emph{Designing a tokamak fusion reactor?How does plasma physics fit in?}, Phys. Plasmas \textbf{22}, 070901 (2015).

\bibitem{Wesson:2004} John Wesson, \emph{Tokamaks}, International Series of Monographs on Physics \textbf{118}, Oxford University Press, Oxford, 3rd edition, 2004.

\bibitem{Yamada:2005} H. Yamada, J.H. Harris, A. Dinklage, E. Ascasibar, F. Sano, S. Okamura, J. Talmadge, U. Stroth, A. Kus, S. Murakami, M. Yokoyama, C.D. Beidler, V. Tribaldos, K.Y. Watanabe, and Y. Suzuki, \emph{Characterization of energy confinement in net-current free plasmas using the extended International Stellarator Database}, Nucl. Fusion \textbf{45}, 1684 (2005).

\bibitem{W7-X:Nature} A. Dinklage, C. D. Beidler, P. Helander, G. Fuchert, H. Maassberg, K. Rahbarnia, T. Sunn Pedersen, Y. Turkin, R. C. Wolf, A. Alonso, T. Andreeva, B. Blackwell, S. Bozhenkov, B. Buttensch\"on, A. Czarnecka, F. Effenberg, Y. Feng, J. Geiger, M. Hirsch, U. H\"ofel, M. Jakubowski, T. Klinger, J. Knauer, G. Kocsis, A. Kr\"amer-Flecken, M. Kubkowska, A. Langenberg, H. P. Laqua, N. Marushchenko, A. Moll\'en, U. Neuner, H. Niemann, E. Pasch,
N. Pablant, L. Rudischhauser, H. M. Smith, O. Schmitz, T. Stange, T. Szepesi, G. Weir, T. Windisch, G. A. Wurden, D. Zhang, and the W7-X Team, \emph{Magnetic configuration effects on the Wendelstein 7-X stellarator}, Nature Physics \textbf{14}, 855 (2018).


 \bibitem{LHD:2108} M. Kobayashi, S. Masuzaki, K. Tanaka, T. Tokuzawa, M. Yokoyama, Y. Narushima, I. Yamada, T. Ido, R. Sekiab, and The LHD Experimental Group, \emph{Core plasma confinement during detachment transition with RMP application in LHD}, Nuclear Materials and Energy \textbf{17}, 137 (2018).


\bibitem{DIII-D:2018} J. M. Park, J. R. Ferron, C. T. Holcomb, R. J. Buttery, W. M. Solomon, D. B. Batchelor, W. Elwasif, D. L. Green, K. Kim, O. Meneghini, M. Murakami, and P. B. Snyder, \emph{Integrated modeling of high $\beta_N$ steady state scenario on DIII-D}, Phys. Plasmas \textbf{25}, 012506 (2018).
 
\bibitem{Klinger:2019} T. Klinger, T. Andreeva, S. Bozhenkov, C. Brandt, R. Burhenn, B. Buttenschön, G. Fuchert, B. Geiger, O. Grulke, H.P. Laqua, N. Pablant, K. Rahbarnia, T. Stange, A. von Stechow, N. Tamura, H. Thomsen, Y. Turkin, T. Wegner, I. Abramovic, S. Äkäslompolo, J. Alcuson, P. Aleynikov, K. Aleynikova, A. Ali, A. Alonso, G. Anda, E. Ascasibar, J.P. Bähner, S.G. Baek, M. Balden, J. Baldzuhn, M. Banduch, T. Barbui, W. Behr, C. Beidler, A. Benndorf, C. Biedermann, W. Biel, B. Blackwell, E. Blanco, M. Blatzheim, S. Ballinger, T. Bluhm, D. Böckenhoff, B. Böswirth, L.-G. Böttger, M. Borchardt, V. Borsuk, J. Boscary, H.-S. Bosch, M. Beurskens, R. Brakel, H. Brand, T. Bräuer, H. Braune, S. Brezinsek, K.-J. Brunner, R. Bussiahn, V. Bykov, J. Cai, I. Calvo, B. Cannas, A. Cappa, A. Carls, D. Carralero, L. Carraro, B. Carvalho, F. Castejon, A. Charl, N. Chaudhary, D. Chauvin, F. Chernyshev, M. Cianciosa, R. Citarella, G. Claps, J. Coenen, M. Cole, M.J. Cole, F. Cordella, G. Cseh, A. Czarnecka, K. Czerski, M. Czerwinski, G. Czymek, A. da Molin, A. da Silva, H. Damm, A. de la Pena, S. Degenkolbe, C.P. Dhard, M. Dibon, A. Dinklage, T. Dittmar, M. Drevlak, P. Drewelow, P. Drews, F. Durodie, E. Edlund, P. van Eeten, F. Effenberg, G. Ehrke, S. Elgeti, M. Endler, D. Ennis, H. Esteban, T. Estrada, J. Fellinger, Y. Feng, E. Flom, H. Fernandes, W.H. Fietz, W. Figacz, J. Fontdecaba, O. Ford, T. Fornal, H. Frerichs, A. Freund, T. Funaba, A. Galkowski, G. Gantenbein, Y. Gao, J. García Regaña, D. Gates, J. Geiger, V. Giannella, A. Gogoleva, B. Goncalves, A. Goriaev, D. Gradic, M. Grahl, J. Green, H. Greuner, A. Grosman, H. Grote, M. Gruca, C. Guerard, P. Hacker, X. Han, J.H. Harris, D. Hartmann, D. Hathiramani, B. Hein, B. Heinemann, P. Helander, S. Henneberg, M. Henkel, J. Hernandez Sanchez, C. Hidalgo, M. Hirsch, K.P. Hollfeld, U. Höfel, A. Hölting, D. Höschen, M. Houry, J. Howard, X. Huang, Z. Huang, M. Hubeny, M. Huber, H. Hunger, K. Ida, T. Ilkei, S. Illy, B. Israeli, S. Jablonski, M. Jakubowski, J. Jelonnek, H. Jenzsch, T. Jesche, M. Jia, P. Junghanns, J. Kacmarczyk, J.-P. Kallmeyer, U. Kamionka, H. Kasahara, W. Kasparek, Y.O. Kazakov, N. Kenmochi, C. Killer, A. Kirschner, R. Kleiber, J. Knauer, M. Knaup, A. Knieps, T. Kobarg, G. Kocsis, F. Köchl, Y. Kolesnichenko, A. Könies, R. König, P. Kornejew, J.-P. Koschinsky, F. Köster, M. Krämer, R. Krampitz, A. Krämer-Flecken, N. Krawczyk, T. Kremeyer, J. Krom, M. Krychowiak, I. Ksiazek, M. Kubkowska, G. Kühner, T. Kurki-Suonio, P.A. Kurz, S. Kwak, M. Landreman, P. Lang, R. Lang, A. Langenberg, S. Langish, H. Laqua, R. Laube, S. Lazerson, C. Lechte, M. Lennartz, W. Leonhardt, C. Li, C. Li, Y. Li, Y. Liang, C. Linsmeier, S. Liu, J.-F. Lobsien, D. Loesser, J. L. Cisquella, J. Lore, A. Lorenz, M. Losert, A. Lücke, A. Lumsdaine, V. Lutsenko, H. Maaßberg, O. Marchuk, J.H. Matthew, S. Marsen, M. Marushchenko, S. Masuzaki, D. Maurer, M. Mayer, K. McCarthy, P. McNeely, A. Meier, D. Mellein, B. Mendelevitch, P. Mertens, D. Mikkelsen, A. Mishchenko, B. Missal, J. Mittelstaedt, T. Mizuuchi, A. Mollen, V. Moncada, T. Mönnich, T. Morisaki, D. Moseev, S. Murakami, G. Náfrádi, M. Nagel, D. Naujoks, H. Neilson, R. Neu, O. Neubauer, U. Neuner, T. Ngo, D. Nicolai, S.K. Nielsen, H. Niemann, T. Nishizawa, R. Nocentini, C. N\"uhrenberg, J. N\"uhrenberg, S. Obermayer, G. Offermanns, K. Ogawa, J. Ölmanns, J. Ongena, J.W. Oosterbeek, G. Orozco, M. Otte, L. Pacios Rodriguez, N. Panadero, N. Panadero Alvarez, D. Papenfuß, S. Paqay, E. Pasch, A. Pavone, E. Pawelec, T.S. Pedersen, G. Pelka, V. Perseo, B. Peterson, D. Pilopp, S. Pingel, F. Pisano, B. Plaum, G. Plunk, P. Pölöskei, M. Porkolab, J. Proll, M.-E. Puiatti, A. Puig Sitjes, F. Purps, M. Rack, S. Récsei, A. Reiman, F. Reimold, D. Reiter, F. Remppel, S. Renard, R. Riedl, J. Riemann, K. Risse, V. Rohde, H. Röhlinger, M. Romé, D. Rondeshagen, P. Rong, B. Roth, L. Rudischhauser, K. Rummel, T. Rummel, A. Runov, N. Rust, L. Ryc, S. Ryosuke, R. Sakamoto, M. Salewski, A. Samartsev, E. Sanchez, F. Sano, S. Satake, J. Schacht, G. Satheeswaran, F. Schauer, T. Scherer, J. Schilling, A. Schlaich, G. Schlisio, F. Schluck, K.-H. Schl\"uter, J. Schmitt, H. Schmitz, O. Schmitz, S. Schmuck, M. Schneider, W. Schneider, P. Scholz, R. Schrittwieser, M. Schr\"oder, T. Schr\"oder, R. Schroeder, H. Schumacher, B. Schweer, E. Scott, S. Sereda, B. Shanahan, M. Sibilia, P. Sinha, S. Sipliä, C. Slaby, M. Sleczka, H. Smith, W. Spiess, D.A. Spong, A. Spring, R. Stadler, M. Stejner, L. Stephey, U. Stridde, C. Suzuki, J. Svensson, V. Szabó, T. Szabolics, T. Szepesi, Z. Szökefalvi-Nagy, A. Tancetti, J. Terry, J. Thomas, M. Thumm, J.M. Travere, P. Traverso, J. Tretter, H. Trimino Mora, H. Tsuchiya, T. Tsujimura, S. Tulipán, B. Unterberg, I. Vakulchyk, S. Valet, L. Vano, B. van Milligen, A.J. van Vuuren, L. Vela, J.-L. Velasco, M. Vergote, M. Vervier, N. Vianello, H. Viebke, R. Vilbrandt, A. Vorköper, S. Wadle, F. Wagner, E. Wang, N. Wang, Z. Wang, F. Warmer, T. Wauters, L. Wegener, J. Weggen, Y. Wei, G. Weir, J. Wendorf, U. Wenzel, A. Werner, A. White, B. Wiegel, F. Wilde, T. Windisch, M. Winkler, A. Winter, V. Winters, S. Wolf, R.C. Wolf, A. Wright, G. Wurden, P. Xanthopoulos, H. Yamada, I. Yamada, R. Yasuhara, M. Yokoyama, M. Zanini, M. Zarnstorff, A. Zeitler, D. Zhang, H. Zhang, J. Zhu, M. Zilker, A. Zocco, S. Zoletnik and M. Zuin, \emph{Overview of first Wendelstein 7-X high-performance operation}, Nucl. Fusion \textbf{59}, 112004 (2019).




\bibitem{Green:2003} B. J. Green for the ITER International Team and Participant Teams, \emph{ITER: burning plasma physics experiment}
 Plasma Phys. Control. Fusion \textbf{45},  687 (2003).






\end{thebibliography}
\end{document}